\documentclass[floatfix,aps,twocolumn,showpacs,superscriptaddress,amsmath,amstex,
amssymb,longbibliography, nofootinbib]{revtex4-2}

\usepackage[autostyle=true]{csquotes}

\usepackage{times}
\usepackage{xcolor}
\usepackage{mathtools}
\usepackage{textcomp}
\usepackage{gensymb}
\usepackage{graphicx}
\usepackage{siunitx}
\usepackage{ulem}
\usepackage{physics}
\usepackage{appendix}
\usepackage{verbatim}
\usepackage{mathrsfs}
\usepackage{amsmath}
\usepackage{upgreek}
\usepackage{tikz}
\usetikzlibrary{matrix,calc}
\usepackage{mathdots}
\usepackage{leftindex}
\usepackage[unicode=true,
 bookmarks=true,bookmarksnumbered=true,bookmarksopen=true,bookmarksopenlevel=2,
 breaklinks=false,pdfborder={0 0 1},backref=false,colorlinks=true]
 {hyperref}

\DeclareFixedFont{\ttb}{T1}{txtt}{bx}{n}{10} 
\DeclareFixedFont{\ttbft}{T1}{txtt}{bx}{n}{8} 
\DeclareFixedFont{\ttm}{T1}{txtt}{m}{n}{10}  

\usepackage{color}
\definecolor{deepblue}{rgb}{0,0,0.5}
\definecolor{deepred}{rgb}{0.6,0,0}
\definecolor{deepgreen}{rgb}{0,0.5,0}

\usepackage{listings}

\newcommand\pythonstyle{\lstset{
language=Python,
numbers=left,
numberstyle=\tiny,
basicstyle=\ttfamily\footnotesize,
morekeywords={self},              
keywordstyle=\ttfamily\color{deepblue},
emph={MyClass,_init_},          
emphstyle=\ttb\color{deepred},    
stringstyle=\color{deepgreen},
frame=tb,                         
showstringspaces=false
}}

\lstnewenvironment{python}[1][]
{
\pythonstyle
\lstset{#1}
}
{}

\newcommand\pythoninline[1]{{\pythonstyle\lstinline!#1!}}

 \hypersetup{linkcolor=blue, citecolor=blue, urlcolor=blue, filecolor=blue, pdfpagelayout=OneColumn,
  pdfnewwindow=true, pdfstartview=XYZ, plainpages=false}

\usepackage[all]{hypcap}
\usepackage{color}

\makeatletter
\def\mathcolor#1#{\@mathcolor{#1}}%
\def\@mathcolor#1#2#3{%
  \protect\leavevmode%
  \begingroup\color#1{#2}#3\endgroup%
}%
\newcommand{\msout}[1]{\text{\sout{\ensuremath{#1}}}}%

\newcommand{\sam}[2]{%
\ifmmode%
  \msout{#1}\mathcolor{red}{#2}%
\else%
  \sout{#1}\textcolor{red}{#2}%
\fi}
\newcommand{\samO}[2]{%
\ifmmode%
  \msout{#1}\mathcolor{magenta}{#2}%
\else%
  \sout{#1}\textcolor{magenta}{#2}%
\fi}

\newcommand{\cyr}[2]{%
\ifmmode%
  \msout{#1}\mathcolor{red}{#2}%
\else%
  \sout{#1}\textcolor{red}{#2}%
\fi}

\newcommand{\gab}[2]{%
\ifmmode%
  \msout{#1}\mathcolor{green}{#2}%
\else%
  \sout{#1}\textcolor{green}{#2}%
\fi}

\makeatother

\makeatletter

\newcommand{\vlad}[1]{\textcolor{cyan}{#1}}

\newcommand\Autoref[1]{\@first@ref#1,@}
\def\@throw@dot#1.#2@{#1}
\def\@set@refname#1{
    \edef\@tmp{\getrefbykeydefault{#1}{anchor}{}}%
    \xdef\@tmp{\expandafter\@throw@dot\@tmp.@}%
    \ltx@IfUndefined{\@tmp autorefnameplural}%
         {\def\@refname{\@nameuse{\@tmp autorefname}s}}%
         {\def\@refname{\@nameuse{\@tmp autorefnameplural}}}%
}
\def\@first@ref#1,#2{%
  \ifx#2@\autoref{#1}\let\@nextref\@gobble
  \else%
    \@set@refname{#1}
    \@refname~\ref{#1}
    \let\@nextref\@next@ref
  \fi%
  \@nextref#2%
}
\def\@next@ref#1,#2{%
   \ifx#2@ and~\ref{#1}\let\@nextref\@gobble
   \else, \ref{#1}
   \fi%
   \@nextref#2%
}

\makeatother

\begin{document}

\def\be{\begin{equation}}
\def\ee{\end{equation}}

\def\myvec#1{{\bf #1}}
\def\Esw{\myvec{E}_{sw}}
\def\Hsw{\myvec{H}_{sw}}

\title{
$\texttt{Symdyn}$: an automated algebraic solution for high-order quantum systems
}

\author{D. Mart\'inez-Tibaduiza}
\email{Correspondence to: danielmartinezt@gmail.com}
\affiliation{Instituto de F\'isica, Universidade Federal Fluminense, Avenida Litor\^{a}nea, 24210-346 Niteroi, RJ, Brazil}
\author{Vladimir~Vargas-Calderón}
\affiliation{D-Wave Systems Inc., Burnaby, British Columbia, Canada}
\author{J.~G.~Dueñas}
\affiliation{Information Analysis Group, Special Jurisdiction for Peace, Cra 7 No 63-44, Bogotá, Colombia}
\author{J. ~Fl\'orez-Jim\'enez}
\affiliation{Lyric, Sunnyvale, California 94087, US}
\author{A.~Z.~Khoury}
\affiliation{Instituto de F\'isica, Universidade Federal Fluminense, Avenida Litor\^{a}nea, 24210-346 Niteroi, RJ, Brazil}


\begin{abstract}

Many significant quantum physical systems are characterized by Hamiltonians expressible as a linear combination of time-independent generators of a closed Lie algebra, $\hat{H}(t)=\sum_{l=1}^{L}\eta_{l}(t)\hat{g}_{l}$. The Wei-Norman method provides a framework for determining the coefficients of the corresponding time evolution operator in its factorized representation, $\hat{U}(t) = \prod_{l=1}^{L} e^{ \Lambda_{l}(t)\hat{g}_{l}}$. This work introduces \verb|symdyn|, a Python library that automates the application of this method. The library efficiently computes similarity transformations and the nonlinear differential equations intrinsic to derive Baker-Campbell-Hausdorff-like relations and the time evolution of  high-order quantum systems ($L\geq 6$). We demonstrate its robustness by deriving the time evolution operator for a system of two time-dependent coupled harmonic oscillators. Additionally, we specialize the library to the Lie group \textit{SU}$(N)$, showing its versatility with \textit{SU}$(2)$, \textit{SU}$(3)$ and \textit{SU}$(4)$ examples, relevant to quantum computing. 

\end{abstract}

\maketitle

\section{Introduction}\label{intro}

The state vector of a quantum system $\left|\psi(t)\right\rangle$ evolves according to the Schr\"odinger equation 
$i\frac{d}{d t}\left|\psi(t)\right\rangle=\hat{H}(t)\left|\psi(t)\right\rangle$ \cite{Sakurai-Book-2014}, where $\hbar = 1$ is assumed. The corresponding time evolution operator (TEO) relates the initial state $\left|\psi(t_{o})\right\rangle$ with the state at any time latter $\left|\psi(t)\right\rangle = \hat{U}(t,t_{o})\left|\psi(t_{o})\right\rangle$. The TEO fulfills the initial condition $ \hat{U}(t\rightarrow t_{o},t_{o}) =\hat{1\!\! 1}$, where $\hat{1\!\! 1}$ is the identity, and obeys the differential equation
\begin{equation}
i\frac{d}{d t}\hat{U}(t,t_{o})=\hat{H}(t)\hat{U}(t,t_{o}) \, .
\label{eq:ScroTEO}
\end{equation}
If the Hamiltonian can be expressed as a linear combination of time-independent (and linearly independent) generators $\{\hat{g}_l\}_{l=1}^L$ of a closed Lie algebra    
\begin{equation}
\hat{H}(t)=\sum_{l=1}^{L}\eta_{l}(t)\hat{g}_{l} \, ,
\label{eq:Lieop}
\end{equation}
where $L$ is the algebraic order  
and $\eta_{l}\in\mathbb{C}$ are time-dependent coefficients, then, the TEO is an element of the associated Lie group. 
It is often written in two main representations. The first one, due to Magnus \cite{Magnus_1954, Blanes_2009}, is the unfactorized representation
\begin{align}
\hat{U}(t) =  \exp{\sum_{l=1}^{L}\lambda_{l}(t)\hat{g}_{l}} \, , 
\label{eq:unfacTEO}
\end{align}
where $\lambda_{l}(t)\in\mathbb{C}$.   
From now on, we suppress the initial time argument to simplify the notation.
The second one is the factorized representation, expressed as a product of exponentials 
\begin{align}
\hat{U}(t) = \prod_{l=1}^{L} e^{ \Lambda_{l}(t)\hat{g}_{l}} \, ,
\label{eq:TEONcomp}
\end{align}
where $\Lambda_{l}\in\mathbb{C}$. Notice that there are $L!$ different--but equivalent in principle-- ways of ordering the exponentials in the expression above, where each arrangement has a different set of coefficients. Note also that the algebra generators need not be Hermitian. 
In the 1960s \cite{Wei_1963, Wei_1964}, Wei and Norman developed this representation which has proven to be particularly useful for computing the state vector of quantum systems and for allowing a simple interpretation of the effect that each component of the TEO, corresponding to the exponential of a given generator, has on specific states. 

The process of obtaining the coefficients $\Lambda_{l}$ from the coefficients of the Hamiltonian $\eta_{l}$ is known as the The Wei-Norman method (WNM) and results in the exact solution of the TEO. On the other hand, obtaining them from the coefficients $\lambda_{l}$ of the unfactorized representation is known as the Wei-Norman factorization, resulting in the so-called BCH-like relations (BCH from Baker–Campbell–Hausdorff). For a given algebra, in both processes a very similar system of coupled nonlinear differential equations arises that relate both sets of coefficients. The main difference is that the equations are time-dependent in the WNM while they are time-independent in the Wei-Norman factorization. Moreover, its structure and solution 
depends on the choice of the specific set of generators used to expand the algebra, \textit{i.e.} the basis, and the ordering of the exponentials in the factorization.

While the factorized representation is globally valid for all solvable Lie algebras, there is no such guarantee for semisimple Lie algebras, where finding an optimal basis to simplify the problem is often non-trivial. Fortunately, for classical simple Lie algebras, which are relevant to this work, a Cartan–Weyl basis yields a globally valid solution where the system of coupled differential equations is reduced to a hierarchy of block-decoupled matrix Riccati equations \cite{Charzy_ski_2013, Charzy_ski_2015}. Due to its robustness, the WNM has gained importance in fields such as optimal control \cite{Brockett_1972, Brockett_1973, Bloch_2015} and quantum physics \cite{Wilcox-1967, Dell_Anno_2006, Carinena_2015, Kosloff_2017, Qvarfort_2021, Duriez_2022}, including open-system dynamics  \cite{Qvarfort_2025}.

Despite these advantages, applying the WNM to high-order ($L \geq 6$  \cite{Vsnob_2017}) quantum systems has remained challenging, mainly from the rapid growth of the algebraic order with increasing group dimension. For example, the special unitary group \textit{SU}$(N)$ has $L=N^{2}-1$ generators in its associated algebra $\mathfrak{su}(N)$. Consequently, there is a substantial increase in the number of coupled nonlinear differential equations needed to determine the TEO, which at some point become intractable to derive by hand. 
Moreover, if the TEO needs to be computed with a different order of exponentials or a different basis, the calculations must be redone from scratch.

To overcome these and other limitations involving high-dimensional applications, we introduce \verb|symdyn|, a Python library that automates the WNM. To our knowledge, no open-source software exists for this purpose. The library also computes nested commutators (Lie products) and nested similarity transformations (BCH relations), enabling the calculation of Hamiltonian evolution or any smooth function of the generators in the Heisenberg picture.

This paper is organized as follows: in Section \ref{WNM_algorithm}, we detail the WNM, highlighting key results for the automation process. We focus on the calculation of the derivative of a Lie group element, which is fundamental to the WNM and the Wei-Norman factorization, and examine the impact of basis choice. 
Meanwhile, we analytically recover non-trivial results for quantum systems associated with the low-order Lie algebras $\mathfrak{su}(1,1)$, $\mathfrak{su}(2)$, $\mathfrak{sl}(2)$ and $\mathfrak{so}(2,1)$  \cite{Teuber_2020, DMT-PRA-2023}, relevant for systems such as the time-dependent harmonic oscillator and qubit dynamics.
We conclude with the derivation of two \textit{SU}$(2)$ gates essential in a configuration for universal quantum computing. 
In Section \ref{symdyn}, we demonstrate \verb|symdyn|'s efficiency for high-order quantum systems by deriving the TEO differential equations for a system of two coupled time-dependent harmonic oscillators with time-dependent coupling \cite{Ramos_2021}. This example is used to provide an overview of the usage of \verb|symdyn|. 
In Section \ref{SUNsec} we specialize our results to  \textit{SU}$(N)$ by providing a generic Cartan-Weyl basis for the associated algebra, allowing us to use the library for arbitrary $N$. We then apply our results to \textit{SU}$(3)$ and \textit{SU}$(4)$, and complete the description of quantum gates for universal quantum computation. 
We close with conclusions and perspectives in 
Section \ref{CONclu}.

%

\section{Wei-Norman method as an algorithm}\label{WNM_algorithm}

Here, we provide a comprehensive and detailed breakdown of the WNM highlighting key results for its automation. For a recent tutorial of the method, including phase space description, open-system dynamics, and applications in quantum optics, see Ref. \cite{Qvarfort_2025}. 

For simplicity, we will refer to any linear combination as defined in Eq.  (\ref{eq:Lieop}) as a \textit{Lie vector}, and quantum systems characterized by Hamiltonians that are Lie vectors as \textit{quantum Lie systems} (QLSs)  with order given by the number of generators. 
To make the calculation of the derivative of Eq.  (\ref{eq:TEONcomp}), which is fundamental for the WNM, more intuitive, we introduce a graphical representation of the structure constants in a tensor arrangement. To illustrate the capabilities  of the WNM, we analytically obtain similarity transformations, BCH-like relations, and the differential equations for the TEO coefficients corresponding to $\mathfrak{su}(1,1)$, $\mathfrak{su}(2)$, $\mathfrak{sl}(2)$ and $\mathfrak{so}(2,1)$ QLSs. Then, we use $\mathfrak{su}(2)$ to discuss decoupling in an appropriated basis. Further, we show how the formalism allows us to build the Hadamard and $T$ gates. 

We would like to point out that the goal of the WNM is to find the TEO when dealing with Hamiltonians with $L \geq 2$ satisfying $[\hat{H}(t),\hat{H}(t')]\neq 0$. The case $L=1$ implies that Eqs. (\ref{eq:unfacTEO}) and (\ref{eq:TEONcomp}) are  equal, so $[\hat{H}(t),\hat{H}(t')]=0$ and there is nothing to factorize. Moreover, for the most general Hamiltonian ($L \geq 2$) commuting with itself at different  times, the computation of the TEO is straightforward \cite{Sakurai-Book-2014}%
\begin{align}
\hat{U}(t) =  \exp{-i \int_{0}^{t}dt' \hat{H}(t')} 
\, , 
\label{eq:unfacTEOsol}
\end{align}
where we have set $t_{o}=0$.


\subsection{Derivative of a factorized Lie group element} \label{derivative}

At the heart of the WNM is the fact that the derivative of an element of a Lie group in its factorized representation can be written as a Lie vector multiplied by the element itself. We show below that to obtain this result it is necessary to compute the set of all possible similarity transformations between the generators. Furthermore, we show how these results can be obtained systematically.

Without any loss of generality, let us consider for the product in Eq. (\ref{eq:TEONcomp}) the ordering
\begin{align}
\hat{U}= e^{\Lambda_{1}\hat{g}_{1}}e^{\Lambda_{2}\hat{g}_{2}}...e^{\Lambda_{L}\hat{g}_{L}}\, ,
\label{eq:TEONcompordered}
\end{align}
where we have omitted the temporal dependence in the argument of the coefficients for simplicity of notation. We shall do that along the text whenever there is no risk of confusion. 
Using the product rule, and that  $e^{\theta\hat{A}}e^{-\theta\hat{A}} =e^{-\theta\hat{A}}e^{\theta\hat{A}} =\hat{1\!\! 1}$ for arbitrary $\hat{A}$ and $\theta\in\mathbb{C}$, it is straightforward to show that \begin{footnotesize}
\begin{align}
\frac{d}{d t}\hat{U} =&  \left(\dot{\Lambda}_{1}e^{\Lambda_{0}\hat{g}_{0}}\hat{g}_{1}e^{-\Lambda_{0}\hat{g}_{0}}+\dot{\Lambda}_{2} e^{\Lambda_{0}\hat{g}_{0}}e^{\Lambda_{1}\hat{g}_{1}}\hat{g}_{2} e^{-\Lambda_{1}\hat{g}_{1}}e^{-\Lambda_{0}\hat{g}_{0}}+\ldots \right.\nonumber\\[3pt]
&\left. + \dot{\Lambda}_{L} e^{\Lambda_{0}\hat{g}_{0}}\cdots e^{\Lambda_{L-1}\hat{g}_{L-1}}\hat{g}_{L} e^{-\Lambda_{L-1}\hat{g}_{L-1}}\cdots e^{-\Lambda_{0}\hat{g}_{0}} \right)\hat{U} \,,
\label{eq:tiderTEONcomp}
\end{align}
\end{footnotesize}where we defined $\Lambda_{0}\equiv 1$ and 
$\hat{g}_{0}\equiv\hat{1\!\! 1}$, and the overdot indicates time derivative. Notice that, 
depending on the algebra, one of the generators 
may be the identity again. 
We next show that the above sum of nested similarity transformations, inside the parentheses, is a Lie vector and calculate it explicitly. 


 \subsubsection{Single similarity transformations}
 \label{Singsim}

Initially, to calculate a \textit{single similarity transformation} we can use the well-known BCH lemma\footnote{The original BCH formula can be straightforwardly deduced using the BCH lemma as 
$e^{\hat{g}_{i}}e^{\hat{g}_{j}}=\exp[\theta\hat{g}_{i}+\int_{0}^{\theta}{d\theta' e^{\theta'\hat{g}_{i}}\hat{g}_{j}e^{-\theta'\hat{g}_{i}}}] |_{\theta=1}\,$.}, which for two arbitrary generators reads \cite{Sakurai-Book-2014}
\begin{align}
e^{\Lambda_{i}\hat{g}_{i}}\hat{g}_{j}\, e^{-\Lambda_{i}\hat{g}_{i}} =&\, \hat{g}_{j}+\Lambda_{i}\left[\hat{g}_{i},\hat{g}_{j}\right]+\frac{(\Lambda_{i})^{2}}{2!}\left[\hat{g}_{i},\left[\hat{g}_{i},\hat{g}_{j}\right]\right]+ \nonumber \\[3pt]
& +\frac{(\Lambda_{i})^{3}}{3!}\left[\hat{g}_{i},\left[\hat{g}_{i},\left[\hat{g}_{i},\hat{g}_{j}\right]\right]\right]+...\,\,. 
\label{BCHkey}
\end{align}
The nested Lie products above correspond to derivatives of the similarity transformation in Eq. (\ref{BCHkey}) with respect to $\Lambda_{i}$, subsequently evaluated at $\Lambda_{i}=0$ \cite{Barnett-1997}. More specifically, if we introduce the linear operator $\text{ad}\hat{g}_{i}$ as 
\begin{equation}
(\text{ad}\hat{g}_{i})^{n} \hat{g}_{j}\equiv
\frac{d^{n} (e^{\Lambda_{i}\hat{g}_{i}}\hat{g}_{j}\, e^{-\Lambda_{i}\hat{g}_{i}})}{d (\Lambda_{i})^{n}}\Big|_{\Lambda_{i}=0} \, ,
\label{eq:adjugate}
\end{equation}
%
with $(\text{ad}\hat{g}_{i})^{0}\hat{g}_{j}\equiv\hat{g}_{j}$, then we can write a single Lie product as 
\begin{equation}
(\text{ad}\hat{g}_{i})\hat{g}_{j}=[\hat{g}_{i},\hat{g}_{j}] \, , 
\label{eq:adop}
\end{equation}
and the powers of this operator will indicate the number of Lie products that are nested. For instance, 
\begin{equation}
(\text{ad}\hat{g}_{i})^{3}\hat{g}_{j} =[\hat{g}_{i},[\hat{g}_{i},[\hat{g}_{i},\hat{g}_{j}]]] \,. 
\label{eq:adoppow}
\end{equation}
With this definition, the right hand side of Eq. (\ref{BCHkey}) is written more compactly,
\begin{align}
e^{\Lambda_{i}\hat{g}_{i}}\hat{g}_{j}\, e^{-\Lambda_{i}\hat{g}_{i}} = (e^{\Lambda_{i}\text{ad}\hat{g}_{i}})\hat{g}_{j} \,.  
\label{BCHkey2}
\end{align}

\begin{figure}[h!]
\includegraphics*[width=6.0cm]{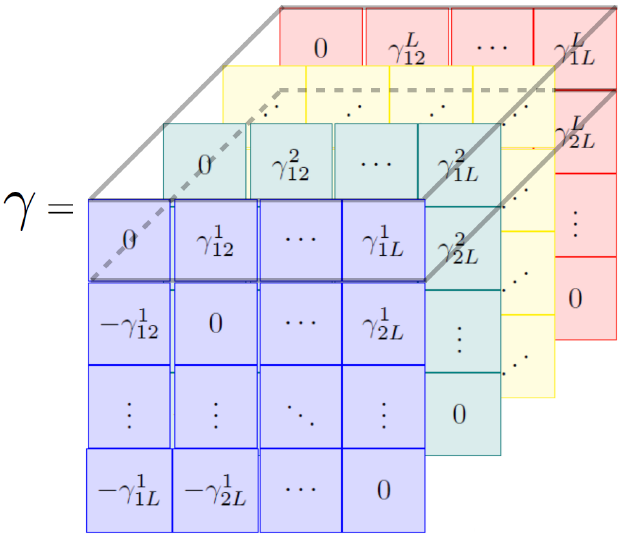}
\caption{Representation of the Structure Tensor $\gamma$ and its top transverse matrix 
$\boldsymbol{\Upsilon}^{1}$.}
\label{fig:LieTensorrow}
\end{figure}

For closed Lie algebras, the Lie product of any two generators is a Lie vector, namely,
\begin{equation}
(\text{ad}\hat{g}_{i})\hat{g}_{j}=\sum_{l=1}^{L}\gamma_{i j}^{l}\hat{g}_{l}  \, ,
\label{eq:Lieproser}
\end{equation}
where the scalars $\gamma_{i j}^{l}$ are called the structure constants, satisfying $\gamma_{i j}^{l}=-\gamma_{j i}^{l}$, $\gamma_{i j}^{l}=0$ for $i= j$, and $\gamma_{0 j}^{l}=\gamma_{i 0}^{l}=0$, as the identity commutes with all the elements of the group. Consequently, nested Lie products can be written in terms of the structure constants as
\begin{align}
(\text{ad}\hat{g}_{i})^{2}\hat{g}_{j}&=\sum_{l=1}^{L}\sum_{s_{1}=1}^{L}\gamma_{i j}^{s_{1}}\gamma_{i s_{1}}^{l}\hat{g}_{l}  \, , \nonumber\\[3pt]
(\text{ad}\hat{g}_{i})^{3}\hat{g}_{j}&=\sum_{l=1}^{L}\sum_{s_{1}=1}^{L}\sum_{s_{2}=1}^{L}\gamma_{i j}^{s_{1}}\gamma_{i s_{1}}^{s_{2}}\gamma_{i s_{2}}^{l}\hat{g}_{l} \,,\,  
\label{eq:Lieproserser}
\end{align}
and so on. As it can be noted, they are also Lie vectors. To help with their calculation, we introduce the  \textit{Structure Tensor} $\gamma$, whose components are the structure constants $\gamma_{i j}^{l}$. The two covariant components identify the generators  involved in the Lie product, and the contravariant component indicates which generator of the resulting Lie vector it is the coefficient of. This tensor can be viewed as the ordered set of all Lie vectors resulting from the Lie products between all the generators of the algebra, and can be represented as in Fig. \ref{fig:LieTensorrow}, where each vertical layer is a skew-symmetric matrix corresponding to $l=1,2,...,L$, with $i$ indexing its rows and $j$ its columns. 
%
Thus, $\gamma$ contains all the relevant information about the algebra for a particular ordering of the exponential generators in the factorized representation. Changing this ordering involves reordering the entries of $\gamma$. 

Each value of $i=1,2,...,L\,$ determines an  $L\times L$ (transverse) matrix in $\gamma$, formed by the set of Lie vectors resulting from the Lie product between the $i$-th generator and each of the generators of the algebra, including itself. We define the set of these matrices as $\lbrace\boldsymbol{\Upsilon}\rbrace_{i=1}^L$, and choose, for future convenience, to index their rows by $j$ and their columns by $l$. This is a slightly different but equivalent definition to that in Ref. \cite{Altafini_2005}.  
The $j$-th row of $\boldsymbol{\Upsilon}^{i}$ contains the coefficients of the Lie vector resulting from the Lie product between the $i$-th and the $j$-th generator of the algebra, and 
therefore is null when $j=i$. 
For instance, the matrix corresponding to $\boldsymbol{\Upsilon}^{1}$ is the one at the top of $\gamma$, as indicated in  Fig. \ref{fig:LieTensorrow}, having its first row null. 
Using this definition it can be inferred from Eqs. (\ref{eq:Lieproserser}) that nested Lie products are equivalent to the powers of $\boldsymbol{\Upsilon}^{i}$ \cite{Gilmore_1974}. 
More specifically \footnote{More generally,  $(\text{ad}\hat{g}_{n})\cdots(\text{ad}\hat{g}_{k})(\text{ad}\hat{g}_{i})\hat{g}_{j}= \sum_{l=1}^{L}\left. (\boldsymbol{\Upsilon}^{n} \ldots \boldsymbol{\Upsilon}^{k}\boldsymbol{\Upsilon}^{i})\right. _{jl}\hat{g}_{l}$. This result could be useful to automate the construction of exact effective Hamiltonians \cite{Chakraborty_2024}.}, 
\begin{align}
(\text{ad}\hat{g}_{i})^{m}\hat{g}_{j}= \sum_{l=1}^{L}\left(  (\boldsymbol{\Upsilon}^{i})^{m}\right)_{jl}\hat{g}_{l} \,.  
\label{eq:gammapower}
\end{align}
This implies that we can write a similarity transformation as
\begin{align}
(e^{\Lambda_{i}\text{ad}\hat{g}_{i}})\hat{g}_{j}&=\sum_{l=1}^{L} b_{j l}^{i}\hat{g}_{l} \,,  
\label{eq:expgral}
\end{align}
where we defined
\begin{align}
b_{j l}^{i} &\equiv \sum_{m=0}^{\infty}\frac{(\Lambda_{i})^{m}}{m!}\left( (\boldsymbol{\Upsilon}^{i})^{m}\right) _{jl} =\left( e^{\Lambda_{i} \boldsymbol{\Upsilon}^{i}}\right)_{jl} \, , 
\label{eq:lambdatens}
\end{align}
for $i, j \geq 1$. We also define  $b_{j l}^{0}\equiv\delta_{j l}$  and 
$(e^{\Lambda_{i}\text{ad}\hat{g}_{i}})\hat{g}_{0}=\hat{g}_{0}$,
where $\delta$ is the Kronecker delta. 
It must be noted that $b_{j l}^{i}$ are the components of a tensor made of all Lie vectors resulting from all possible similarity transformations between the generators of the algebra, with a coefficient multiplying the generator in the exponential. 
We will refer to this tensor as $\mathfrak{b}$, with  component matrices $\textbf{b}^{i}$, so that 
\begin{align}
\textbf{b}^{i} = e^{\Lambda_{i} \boldsymbol{\Upsilon}^{i}} \, , 
\label{eq:lambdatens2}
\end{align}
with $\textbf{b}^{0}$ the $L\times L$ identity  matrix. The calculation of a particular  similarity transformation is now reduced to selecting a row ($j$-index) of a $\textbf{b}$ matrix ($i$-index), and identify its entries ($l$-index) with the components of the resulting Lie vector. We shall call $\mathfrak{b}$ as the BCH tensor, and  $\textbf{b}^{i}$ as the BCH matrices. 

Let us stress the importance of the BCH tensor by calculating it for the Lie algebras $\mathfrak{su}(1,1)$, $\mathfrak{su}(2)$, $\mathfrak{sl}(2)$ and $\mathfrak{so}(2,1)$. There are many prominent physical quantum systems associated with these algebras of order three. For instance, a time-dependent qubit, belonging to the Lie algebra $\mathfrak{su}(2)$ \cite{de-Clercq-2016}, or a time-dependent quantum harmonic oscillator, belonging to the Lie algebra $\mathfrak{su}(1,1)$ \cite{Husimi-1953, DMT-PS-2020, DMT-JPHYSB-2021}, being fundamental systems of great importance  
in many fields of physics \cite{Ahlbrandt_1996,  Li_2009, Kosloff_2017, Bruschi_2018,  Bruschi_2019, Qvarfort_2019, Elouard_2020, Dann_2020, Dupays_2021, Qvarfort_2021, Abah_2022, Mihalcea_2022,  Fei_2022, Benjamin_2022nano}, allowing for the exploration of effects such as squeezing \cite{WALLS-1983, Ma_2011} and spin inversion \cite{Silver_1985, de_Graaf_2019}, both crucial for advancing quantum technologies. Even some models for coupled harmonic oscillators, coupled spins or coupled two-photon lasers can also be associated with these algebras \cite{Shen_2003}. Moreover, they can be parameterized to be worked out within an unified way, as we show in Table \ref{CommRe}. There, the Lie products between the algebra generators are given in Table \ref{CommRe}.a, while parameters $\upsilon$ and $\epsilon$ allow to specify the algebra in Table \ref{CommRe}.b. Table \ref{CommRe}.a is therefore a compact way of writing the Structure Tensor, and this kind of table will be called a \textit{structure table}. 
%
\begin{table}[htb]
\includegraphics*[width=7.5cm]{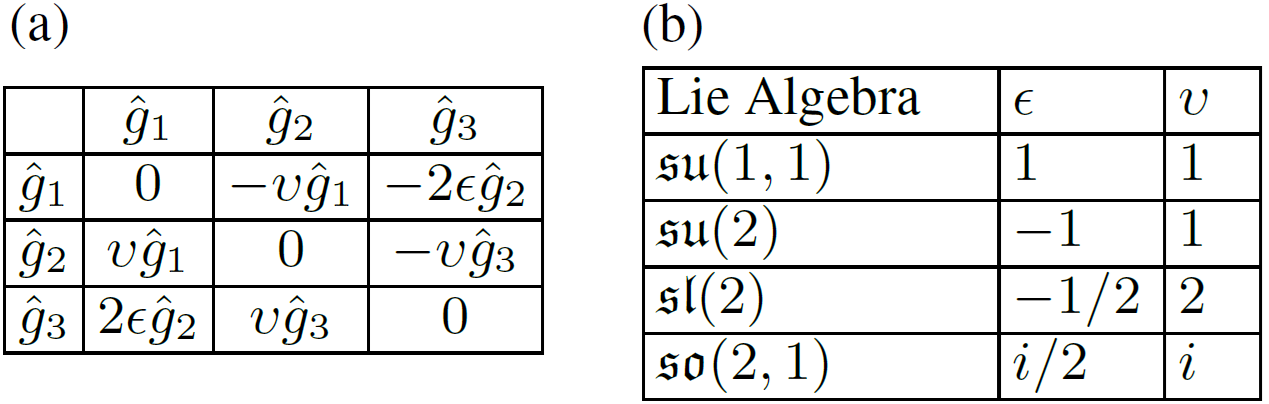}
\caption{(a) Structure table for the Lie algebras $\mathfrak{su}(1,1)$, $\mathfrak{su}(2)$, $\mathfrak{sl}(2)$ and $\mathfrak{so}(2,1)$. Each entry of its inner $3\times 3$ (structure) matrix corresponds to the Lie vector resulting from the ordered Lie product between the generator at the left  with the one at the top. Parameters $\upsilon$ and $\epsilon$ allow to specify the algebra as indicated in (b).}
\label{CommRe}
\end{table}
%
Recall that the transverse matrices $\boldsymbol{\Upsilon}^{i}$ can be rapidly deduced from the Lie products between the generators. Using Table  \ref{CommRe}.a, the transverse matrices are calculated directly to be
\begin{align}
 \boldsymbol{\Upsilon}^{1}=&
 \begin{bmatrix}
    0 & 0 & 0
 \\     
   -\upsilon & 0 &  0 
   \\
        0 & -2\epsilon & 0
\end{bmatrix} \, ,  \,\,\, \boldsymbol{\Upsilon}^{2}=\begin{bmatrix}
    \upsilon & 0 & 0
 \\     
   0 & 0 & 0 
   \\
     0 & 0 & -\upsilon
\end{bmatrix} \, , \nonumber \\
&\,\,\,\,\,\,\,\mbox{and} \,\,\,\,\,\,\,
\boldsymbol{\Upsilon}^{3}=
\begin{bmatrix}
    0 & 2\epsilon  &  0
 \\     
   0 & 0 & \upsilon \\
        0 & 0  & 0
\end{bmatrix}\, . 
\label{eq:LieMatrixsuupsil}
\end{align}
It can be noted that matrices $\boldsymbol{\Upsilon}^{1}$ and $\boldsymbol{\Upsilon}^{3}$ are nilpotent of degree $3$, once  $(\boldsymbol{\Upsilon}^{1})^{3}= (\boldsymbol{\Upsilon}^{3})^{3}=\boldsymbol{0}$, thus to calculate the BCH matrices we only need to compute up to their squares,
\begin{align}
(\boldsymbol{\Upsilon}^{1})^{2}=
 \begin{bmatrix}
    0 & 0 & 0
 \\     
   0 & 0 &  0 
   \\
        2\epsilon\upsilon & 0 & 0
\end{bmatrix} \, ,  \,\,\, (\boldsymbol{\Upsilon}^{3})^{2}=\begin{bmatrix}
    0 & 0 & 2\epsilon\upsilon
 \\     
   0 & 0 & 0 
   \\
     0 & 0 & 0
\end{bmatrix} \, . 
\label{eq:LieMatrixsuuospow}
\end{align}
On the other hand $\boldsymbol{\Upsilon}^{2}$ is diagonal, and then is direct that
\begin{align}
(\boldsymbol{\Upsilon}^{2})^{m}=\begin{bmatrix}
    (\upsilon)^{m} & 0 & 0
 \\     
   0 & 0 & 0 
   \\
     0 & 0 & (-\upsilon)^{m}
\end{bmatrix} \, .
\label{eq:LieMatrixsuuospow}
\end{align}
Finally, for any transverse matrix we have
\begin{align}
(\boldsymbol{\Upsilon}^{i})^{0}=\begin{bmatrix}
    1 & 0 & 0
 \\     
   0 & 1 & 0 
   \\
     0 & 0 & 1
\end{bmatrix} \, .
\label{eq:ident}
\end{align}
Using the above results in Eq. (\ref{eq:lambdatens2}), we can calculate the BCH matrices as
\begin{align}
\textbf{b}^{1} =&  
\begin{bmatrix}
    1 & 0 & 0
 \\     
   -\upsilon\Lambda_{1} & 1 & 0 
   \\
     \epsilon\upsilon(\Lambda_{1})^{2} & 
     -2\epsilon\Lambda_{1} & 1
\end{bmatrix} \, , \,\,\, \textbf{b}^{2} =  \begin{bmatrix}
    e^{\upsilon\Lambda_{2}} & 0 & 0 \nonumber
 \\     
   0 & 1 & 0 
   \\
     0 & 0 & e^{-\upsilon\Lambda_{2}}
\end{bmatrix} \, , \\
&\,\,\,\,\,\,\,\mbox{and} \,\,\,\,\,\,\,\, 
\textbf{b}^{3} =  \begin{bmatrix}
    1 & 2\epsilon\Lambda_{3} & \epsilon\upsilon(\Lambda_{3})^{2}
 \\     
   0 & 1 & \upsilon\Lambda_{3} 
   \\
     0 & 0 & 1
\end{bmatrix} \,.
\label{eq:BCHMatrices3gen}
\end{align}
As mentioned before, any similarity transformation in the algebra can be now easily calculated. For instance, 
\begin{align}
(e^{\Lambda_{1}\text{ad}\hat{g}_{1}})\hat{g}_{2}
&=\sum_{l=1}^{L} b_{2 l}^{1}\hat{g}_{l} \, = -\upsilon\Lambda_{1}\hat{g}_{1} + \hat{g}_{2} \, , 
\label{eq:expgraleplic}
\end{align}
corresponding with the second row of $\textbf{b}^{1}$ in Eq. (\ref{eq:BCHMatrices3gen}). 

It is important to note that the BCH  matrices must be calculated explicitly to proceed with the next steps of the method, and their computation depends on the specific algebra we are working with. 
Fortunately, there are many different techniques and frameworks to calculate matrix exponentials (see for instance \cite{Laufer_1997, Altafini_2005, jax2018github, https://doi.org/10.48550/arxiv.1912.01703}). For our numerical calculations we use SymPy \cite{Meurer_2017}. 
Moreover, by using a suitable basis \cite{Charzy_ski_2013, Charzy_ski_2015}, as we will show in Section \ref{BasPro}, we can expect transverse matrices to be nilpotent or diagonal, simplifying the calculation of their powers as in the above example. 


 \subsubsection{Nested similarity transformations}
 \label{Nestsim}

Now that we have shown how the set of  single similarity transformations can be computed, we can move on to compute \textit{nested similarity transformations}. Using Eq. (\ref{eq:expgral}), it is straightforward to show that $r$ ordered (as in Eq. (\ref{eq:TEONcompordered})) nested similarity transformations can be expressed as
\begin{align}
\prod_{i=0}^{r}(e^{\Lambda_{i}\text{ad}\hat{g}_{i}})\hat{g}_{j}= \sum_{l=1}^{L}\beta_{jl}^{r}\hat{g}_{l} \, , 
\label{eq:sameas}
\end{align}
for $r\in\{1,2,\ldots,L\}$, where we have defined
\begin{align}
\beta_{j l}^{r} &\equiv \sum_{s_{1}=1}^{L}\sum_{s_{2}=1}^{L}\cdots\sum_{s_{r-1}=1}^{L} b_{j s_{1}}^{r} b_{s_{1} s_{2}}^{r-1}\cdots b_{s_{r-2} s_{r-1}}^{2} 
b_{s_{r-1} l}^{1} \nonumber \\  
& = \left( \textbf{b}^{r}\textbf{b}^{r-1}\ldots \textbf{b}^{1} \right)_{j l}
 \, .
\label{eq:betafin}
\end{align}
%
Additionally, any sequence of nested similarity transformations, not necessarily ordered, can be also computed by simply multiplying specific BCH matrices. 


Coming back to Eq. (\ref{eq:tiderTEONcomp}), using  Eq. (\ref{BCHkey2}) it can be put in the form
\begin{align}
\frac{d}{d t}\hat{U} =  \sum_{n=1}^{L}\dot{\Lambda}_{n}\left(\prod_{i=0}^{n-1} (e^{\Lambda_{i} \text{ad}\hat{g}_{i}}) \hat{g}_{n}\right)\hat{U} \, .
\label{eq:tiderTEONcompfin}
\end{align}
The expression within the parentheses with the help of Eq. (\ref{eq:sameas}) becomes
\begin{align}
\prod_{i=0}^{n-1} (e^{\Lambda_{i} \text{ad}\hat{g}_{i}}) \hat{g}_{n}= \sum_{l=1}^{L}\xi_{nl}\hat{g}_{l}\, ,
\label{eq:tiderTEONcompfinap}
\end{align}
where we defined $\xi_{nl}\equiv \beta_{n l}^{n-1}$ as the elements of the matrix $\boldsymbol{\xi}$, namely, 
\begin{align}
\xi_{n l} = \left( \textbf{b}^{n-1}\textbf{b}^{n-2}\ldots \textbf{b}^{1} \right)_{n l}  \, .   
\label{eq:betafinmat}
\end{align}
Notice that the $n$-th row of $\boldsymbol{\xi}$ equals the $n$-th row of the resulting ordered product of the first $n-1$ BCH matrices. We shall call $\boldsymbol{\xi}$ the \textit{coupling matrix}, for reasons that will become clear later.  Using the above results, we can finally write Eq. (\ref{eq:tiderTEONcompfin}) as
\begin{align}
\frac{d}{d t}\hat{U} =  \left(  \sum_{l=1}^{L}\sum_{n=1}^{L}\dot{\Lambda}_{n}\xi_{nl}
\hat{g}_{l}\right)\hat{U} \, ,
\label{eq:TEOderFin}
\end{align}
and thus we have shown that the derivative of a Lie element in the factorized representation is given by an explicit Lie vector (inside the parenthesis)  times the element itself. 

For clarity, let us calculate the coupling matrix for the low-order Lie algebras we have been working with. 
From Eqs. (\ref{eq:BCHMatrices3gen}) and (\ref{eq:betafinmat}), the following results are straightforward: the first row of $\boldsymbol{\xi}$ is the same as the first row of $\textbf{b}^{0}$, that is, the first row of the identity; the second row of $\boldsymbol{\xi}$ is the same as the second row of $\textbf{b}^{1}$; and the third row of $\boldsymbol{\xi}$ is the same as the  third row of the resulting matrix $\textbf{b}^{2}\textbf{b}^{1}$, a product that can be calculated using Eqs. (\ref{eq:BCHMatrices3gen}) as
\begin{align}
\textbf{b}^{2}\textbf{b}^{1} =  \begin{bmatrix}
    e^{\upsilon\Lambda_{2}} & 0 & 0
 \\     
   -\upsilon\Lambda_{1} & 1 & 0 
   \\
     \epsilon\upsilon \frac{(\Lambda_{1})^{2}}{e^{\upsilon\Lambda_{2}}} & -2\epsilon \frac{\Lambda_{1}}{e^{\upsilon\Lambda_{2}}} & \frac{1}{e^{\upsilon\Lambda_{2}}}
\end{bmatrix} \, .
\label{eq:BCHMatricesprod}
\end{align}
Using the above results we finally obtain
\begin{align}
\boldsymbol{\xi} = 
\begin{bmatrix}
    1 & 0 & 0
 \\     
   -\upsilon\Lambda_{1} & 1 & 0 
   \\
    \epsilon\upsilon \frac{(\Lambda_{1})^{2}}{e^{\upsilon\Lambda_{2}}} & -2\epsilon \frac{\Lambda_{1}}{e^{\upsilon\Lambda_{2}}} & \frac{1}{e^{\upsilon\Lambda_{2}}}
\end{bmatrix} \,.
\label{eq:xiMatrix3gen}
\end{align}
Notice that the determinant of the above coupling matrix is $\det\vert\boldsymbol{\xi}\vert = \exp(-\upsilon\Lambda_{2})\,$, and it vanishes in the limit  $\Lambda_{2}\rightarrow\infty$. Next, we will see the importance of calculating the determinant of the coupling matrix and the role of the basis in the usefulness of the method.

\subsection{TEO coefficients and an appropriated basis} \label{coefficients}

In the previous section, we calculated the specific Lie vector that determines the ratio between the derivative of a Lie element in the factorized representation and the element itself.  For QLSs, such a vector must be proportional to the Hamiltonian. More specifically, using Eqs. (\ref{eq:Lieop}) and (\ref{eq:TEOderFin}) in Eq. (\ref{eq:ScroTEO}) we obtain
\begin{equation}
\sum_{l=1}^{L}\eta_{l}\hat{g}_{l}=
i\sum_{l=1}^{L}\sum_{n=1}^{L}\dot{\Lambda}_{n}\xi_{nl}\hat{g}_{l} \, ,
\label{eq:ScroTEOpart}
\end{equation}
and considering that the generators are linearly independent, we must have 
\begin{equation}
\eta_{l}=i\sum_{n=1}^{L}\dot{\Lambda}_{n}\xi_{nl} \, ,
\label{eq:diffeq}
\end{equation}
which is a set of time-dependent coupled differential equations with the initial condition $\Lambda_{n}=0$ for all $n$ at $t=0$. It can be written in matrix form as
\begin{equation}
\overbrace{\begin{bmatrix}
    \eta_{1}(t)     \\
    \eta_{2}(t)      \\
    \vdots		 \\
    \eta_{L}(t)   
\end{bmatrix}}^{
\mbox{$\boldsymbol{\eta}$}
}
=
i\overbrace{\begin{bmatrix}
    \xi_{11} & \xi_{21} & \dots  & \xi_{L1} \\
    \xi_{12} & \xi_{22} & \dots  & \xi_{L2} \\
    \vdots & \vdots & \ddots & \vdots \\
    \xi_{1L} & \xi_{2L} & \dots  & \xi_{LL}
\end{bmatrix}}^{
\mbox{$\boldsymbol{\xi}$}^{T}
}
\overbrace{\begin{bmatrix}
    \dot{\Lambda}_{1}     \\
    \dot{\Lambda}_{2}      \\
    \vdots		 \\
    \dot{\Lambda}_{L}   
\end{bmatrix}}^{
\mbox{$\dot{\boldsymbol{\Lambda}}$}
}\,\,,
\label{eq:diffeqmatr}
\end{equation}
where ${T}$ indicates the transpose and recall that all the explicit time-dependence is in $\boldsymbol{\eta}$. Now it is clear why we called $\boldsymbol{\xi}$ the coupling matrix. The global validity of the above equations is only possible when $\boldsymbol{\xi}$ is invertible for all time, a condition verifiable by calculating its determinant $\det\vert\boldsymbol{\xi}\vert$. Nevertheless, there is always a local solution, as there exist at least a neighborhood of $ t = 0$ where $\boldsymbol{\xi}$ is invertible. This is consequence from $\det\vert\boldsymbol{\xi}\vert$ be an analytic function of the TEO coefficients and $\det\vert\boldsymbol{\xi}(t\rightarrow 0)\vert=1 \,$ \cite{Wei_1964}. 

Considering $\det\vert\boldsymbol{\xi}\vert\neq 0$, we can multiply by the inverse of $\boldsymbol{\xi}^{T}$ from the left in the above equation and get
\begin{equation}
(\boldsymbol{\xi}^{T}) ^{-1} \boldsymbol{\eta}=i \dot{\boldsymbol{\Lambda}} \,,
\label{eq:Soldiffeqmatr}
\end{equation}
which is a system of block-decoupled\footnote{The name ``block-decoupled" signals that some of the resulting differential equations for high-order QLSs will still be coupled, but in blocks with the maximum size of the algebra rank (maximum number of diagonalizable matrices), as we will see in Secs. \ref{couposc} and \ref{SUNsec}.} nonlinear differential equations known as \textit{the exact solution of the TEO}. On the other hand, if there are singularities in the coupling matrix \textit{i.e.}, values of the TEO coefficients for which $\det\vert\boldsymbol{\xi}\vert = 0$, the solution will not be global. In fact, as mentioned by Wei and Norman, to find a global solution  the choice of basis is critical \cite{Wei_1964}. 
In the next subsection, we will return to this important point related to the usefulness of the method. 
For now, as an example with a global solution, let us obtain the TEO equations for the four algebras we have been working on. 

Using Eq. (\ref{eq:xiMatrix3gen}) together with Eq. (\ref{eq:diffeqmatr}), we obtain directly 
\begin{equation}
\begin{bmatrix}
    \eta_{1}(t)     \\
    \eta_{2}(t)      \\
    \eta_{3}(t)   
\end{bmatrix}
=
i\begin{bmatrix}
    1 & -\upsilon\Lambda_{1}   & \epsilon\upsilon\frac{ (\Lambda_{1})^{2}}{e^{\upsilon\Lambda_{2}}} \\
    0 & 1  & -2\epsilon\frac{ \Lambda_{1}}{e^{\upsilon\Lambda_{2}}} \\
    0 & 0   & \frac{ 1}{e^{\upsilon\Lambda_{2}}}
\end{bmatrix}
\begin{bmatrix}
    \dot{\Lambda}_{1}     \\
    \dot{\Lambda}_{2}      \\
    \dot{\Lambda}_{3}   
\end{bmatrix}\,.
\label{eq:diffeqmatrQLS}
\end{equation}
%
As $\det\vert\boldsymbol{\xi}^{T}\vert = \det\vert\boldsymbol{\xi}\vert=\exp(-\upsilon\Lambda_{2})\,$, one can invert $\boldsymbol{\xi}^{T}$ for all times  except when $\Lambda_{2}\rightarrow\infty$. 
Multiplying by $(\boldsymbol{\xi}^{T}) ^{-1}$ from the left in the above equation and then performing the matrix product, we obtain the system of 
differential equations:
\begin{align}
\label{eq:Riccati}
&\dot{\Lambda}_{1}+\epsilon\upsilon(i \eta_{3})(\Lambda_{1})^{2}+ \upsilon(i \eta_{2})\Lambda_{1} +i \eta_{1}=0 \, , \\
\label{eq:sollambdacTD}
&\dot{\Lambda}_{2}+2\epsilon(i \eta_{3})\Lambda_{1}+ i\eta_{2} = 0 \, , \\
\label{eq:solgmabdac}
&\dot{\Lambda}_{3}+ (i\eta_{3}) e^{\upsilon\Lambda_{2}} = 0 \,.
\end{align}
%
Eq. (\ref{eq:Riccati}) represents four families of time-dependent complex Riccati equations, each associated with one of the Lie algebras we are considering. The other two equations are solved by quadrature after solving this previous one. The above results are in agreement with the differential equations obtained for $\mathfrak{su}(1,1)$, $\mathfrak{su}(2)$ and $\mathfrak{so}(2,1)$ QLSs in Ref. \cite{DMT-PRA-2023} (see Eqs. (8)-(10) there, taking into account that they considered $\ln{\Lambda_{2}}$ as the coefficient of $\hat{g}_{2}$), and such for the $\mathfrak{sl}(2)$ QLS established in Ref. \cite{Teuber_2020} (see Eqs. (54)-(56) there). 

Although the Riccati equation has no analytical solution for the general time-dependent case  \cite{Tsai_2010}, the time-independent case does have a solution. In Appendix \ref{appA} we use this solution to obtain BCH-like relations for the above algebras, while showing how the Wei-Norman factorization is carried out. In particular, BCH-like relations of $\mathfrak{su}(2)$ will be used to calculate two important one-qubit gates in Section \ref{OQgates} as a confirmation of our results. Let us now explore the issue of the choice of basis to obtain a global solution.


 \subsubsection{Decoupling in an appropriated basis }
 \label{BasPro}
To illustrate the impact of choosing an appropriate basis on the utility of the WNM, let us apply the method for $\mathfrak{su}(2)$ QLSs but this time using the Pauli matrices as a basis, which is the usual basis for working with qubits. More specifically, we consider the generators: $\underline{\hat{g}}_{j}=\frac{i}{2}\hat{\sigma}_j$ ($j=1,2,3$), where 
\begin{align}
\hat{\sigma}_1 =& 
\begin{bmatrix}
    0 & 1 
   \\
     1 & 0
\end{bmatrix}  , \,\,\, 
\hat{\sigma}_2 =  \begin{bmatrix}
    0 & -i 
   \\
     i & 0
\end{bmatrix} , \,\,\,
\hat{\sigma}_3 =  \begin{bmatrix}
    1 & 0 
   \\
     0 & -1
\end{bmatrix}  \,,
\label{eq:PauliMatrices}
\end{align}
are the Pauli matrices, and the idea is to find the TEO
\begin{align}
\hat{U}= e^{\Theta_{1}\underline{\hat{g}}_{1}}e^{\Theta_{2}\underline{\hat{g}}_{2}}e^{\Theta_{3}\underline{\hat{g}}_{3}}\, ,
\label{eq:TEOsu2paili}
\end{align}
corresponding to a given Hamiltonian in this basis
\begin{align}
   \hat{H} = \varsigma_{1}\underline{\hat{g}}_{1}+  \varsigma_{2}\underline{\hat{g}}_{2}+ \varsigma_{3}\underline{\hat{g}}_{3} \, .
\label{eq:rabiHamiltonian}
\end{align}

The well-known commutation relations for Pauli matrices $[\sigma_{j}, \sigma_{k}]=2i\sum_{l=1}^{3}\varepsilon_{jkl}\sigma_{l}$, where $\varepsilon_{jkl}$ is the Levi-Civita symbol, allow us to build the structure table of the system (see Table \ref{CommRePauli}), and with it the Structure Tensor. 
\begin{table}[htb!]
\centering
\begin{tabular}{|l|c|c|c|}
 \hline
  & $\underline{\hat{g}}_{1}$ & $\underline{\hat{g}}_{2}$ & $\underline{\hat{g}}_{3}$ \\
 \hline
$\underline{\hat{g}}_{1}$  & $0$  & $- \underline{\hat{g}}_{3}$ & $ \underline{\hat{g}}_{2}$ \\
\hline
$\underline{\hat{g}}_{2}$  & $ \underline{\hat{g}}_{3}$  & $0$ & $- \underline{\hat{g}}_{1}$ \\
\hline
$\underline{\hat{g}}_{3}$  & $- \underline{\hat{g}}_{2}$  & $\underline{\hat{g}}_{1}$ & $0$ \\ 
\hline
\end{tabular}   \, . 
\label{CommRePauli}
\caption{Structure table for the Lie algebra $\mathfrak{su}(2)$ in the Pauli basis. }
\end{table}
From the latter, one can calculate the corresponding coupling matrix, which is
\begin{small}
\begin{align}
\boldsymbol{\xi}_{P} = 
\begin{bmatrix}
    1 & 0 & 0
 \\     
	0 & \cos{(\Theta_{1})} & 
	-\sin{(\Theta_{1})} 
   \\
    -\sin{(\Theta_{2})} & \sin{(\Theta_{1})}\cos{(\Theta_{2})} & \cos{(\Theta_{1})}\cos{(\Theta_{2})}
\end{bmatrix} \, .
\label{eq:xiMatrix3genPauli}
\end{align}
\end{small}Its determinant is given by $\det\vert\boldsymbol{\xi}_{P}\vert = \cos{(\Theta_{2})}\,$, so it has singularities at $\Theta_{2}=\pi(n+1/2)\,$, with $n\in\mathbb{Z}$. The coupling matrix (as well as its transpose) is therefore non-invertible for all times, and then this solution is not global. In other words, we cannot obtain a global factorized version of the TEO, as in Eq. (\ref{eq:TEOsu2paili}), using the Pauli matrices as the generators of $\mathfrak{su}(2)$, either starting with an unfactorized version of the TEO, or with a Hamiltonian in this basis. Although the inverse of the coupling matrix can be calculated to be
\begin{small}
\begin{align}
\left. \boldsymbol{\xi}_{P}\right. ^{-1} = 
\begin{bmatrix}
    1 & 0 & 0
 \\     
	\sin{(\Theta_{1})}\tan{(\Theta_{2})} & \cos{(\Theta_{1})} & 
	\sin{(\Theta_{1})}\sec{(\Theta_{2})} 
   \\
    \cos{(\Theta_{1})}\tan{(\Theta_{2})} &  -\sin{(\Theta_{1})} & \cos{(\Theta_{1})}\sec{(\Theta_{2})}
\end{bmatrix} \, ,
\label{eq:xiMatrix3genPauliinv}
\end{align}
\end{small}it does exist for all values of $\Theta_{l}$  except in the mentioned periodic singularities \cite{Altafini_2003}. Besides, using the above equation and Eq. (\ref{eq:Soldiffeqmatr}) we can calculate the corresponding TEO differential equations  \begin{small}
\begin{align}
\label{eq:NLEPaili1}
&\dot{\Theta}_{1}+i\tan{(\Theta_{2})}\left[ \varsigma_{2}\sin{(\Theta_{1})}+\varsigma_{3}\cos{(\Theta_{1})}\right] +i\varsigma_{1}=0 \, , \\
\label{eq:NLEPaili2}
&\dot{\Theta}_{2}+ i\left[\varsigma_{2}\cos{(\Theta_{1})}-\varsigma_{3}\sin{(\Theta_{1})}\right]=0 \, ,\\
\label{eq:NLEPaili3}
&\dot{\Theta}_{3}+i\sec{(\Theta_{2})}\left[ \varsigma_{2}\sin{(\Theta_{1})}+\varsigma_{3}\cos{(\Theta_{1})}\right] =0 \,.
\end{align}
\end{small}which are coupled, and decoupling is not evident. Nevertheless, a simple change of basis, as anticipated by Wei and Norman, can change everything. 
If we define ladder operators $\hat{g}_{1}\equiv\hat{\sigma}_{+}=\frac{1}{2}(\hat{\sigma}_1+i\hat{\sigma}_2)$ and $\hat{g}_{3} \equiv\hat{\sigma}_{-}=\frac{1}{2}(\hat{\sigma}_1-i\hat{\sigma}_2)$, together with $\hat{g}_{2}=\frac{1}{2}\hat{\sigma}_{3}$, then the algebra falls in the commutation relations for $\mathfrak{su}(2)$ given in Table \ref{CommRe}.a, and consequently the TEO can be expressed as 
\begin{align}
\hat{U}= e^{\Lambda_{1}\hat{g}_{1}}e^{\Lambda_{2}\hat{g}_{2}}e^{\Lambda_{3}\hat{g}_{3}}\, ,
\label{eq:TEON4gen}
\end{align}
with $\Lambda_{1}$, $\Lambda_{2}$ and $\Lambda_{3}$ given by  Eqs. (\ref{eq:Riccati})-(\ref{eq:solgmabdac}), with $\epsilon=-1$ and $\upsilon=1$. Further, we can calculate the TEO in terms of the Pauli matrices 
as follows. From their definitions and Eqs. (\ref{eq:PauliMatrices}), it can be shown that $\hat{g}_{1}$ and $\hat{g}_{3}$ are nilpotent of degree 2, and therefore
\begin{align}
e^{\Lambda_{1}\hat{g}_{1}}=\hat{1\!\! 1}+\frac{\Lambda_{1}}{2}(\hat{\sigma}_1+i\hat{\sigma}_2)\, ,
   \,\,e^{\Lambda_{3}\hat{g}_{3}}=\hat{1\!\! 1}+\frac{\Lambda_{3}}{2}(\hat{\sigma}_1-i\hat{\sigma}_2)\, .
\label{eq:Laddcal}
\end{align}
On the other hand 
\begin{align}
e^{\Lambda_{2}\hat{g}_{2}}=&\sum_{n=0}^{\infty}\frac{1}{n!}\left(
\frac{\Lambda_{2}}{2}\right)^{n}(\hat{\sigma}_{3})^{n}  \nonumber \\
=&\cosh{\left(\frac{\Lambda_{2}}{2}\right)}\hat{1\!\! 1}+\sinh{\left(\frac{\Lambda_{2}}{2}\right)} \hat{\sigma}_{3}
\, ,
\label{eq:sigmazcal}
\end{align}
where we have used that $\hat{\sigma}_{i}^{2}=\hat{1\!\! 1}$ for $i=1,2,3$. Substituting Eqs. (\ref{eq:Laddcal}) and (\ref{eq:sigmazcal}) in Eq. (\ref{eq:TEON4gen}), it is straightforward to show that
\begin{align}
\hat{U}=& \frac{e^{-\frac{\Lambda_{2}}{2}}}{2}\left(
(e^{\Lambda_{2}}+1+\Lambda_{1}\Lambda_{3})\hat{1\!\! 1}+(\Lambda_{1}+\Lambda_{3})\hat{\sigma}_{1} + \right. 
 \nonumber \\ 
 &\,\,\,\,\,\,\,\,\,\,\,\,\,\,\,\left. +(\Lambda_{1}-\Lambda_{3})\hat{\sigma}_{2}+(e^{\Lambda_{2}}-1+\Lambda_{1}\Lambda_{3})\hat{\sigma}_{3}\right)\, ,
\label{eq:TEONQubitPauli}
\end{align}
where we have used the well-known property of the Pauli matrices $\hat{\sigma}_{j}\hat{\sigma}_{k}=\delta_{j k} \hat{1\!\! 1}+ i\sum_{l=1}^{3}\varepsilon_{jkl}\hat{\sigma}_{l}$. Now, using Eqs. (\ref{eq:PauliMatrices})
we can compute the TEO in its explicit matrix form as
\begin{align}
\hat{U}=e^{-\frac{\Lambda_{2}}{2}}\begin{bmatrix}
    e^{\Lambda_{2}}+ \Lambda_{1}\Lambda_{3} & \Lambda_{1} 
   \\
     \Lambda_{3} & 1
\end{bmatrix}\, ,
\label{eq:TEONQubit}
\end{align}
which satisfies $\det\vert\hat{U}\vert = 1$. Notice that we could have computed the TEO in the above explicit form by directly substituting the generators in their matrix form into the equation (\ref{eq:TEON4gen}), then computing the matrix exponentials and finally doing their products.
This direct way of computing the 
TEO will be used when working with higher-order 
algebras.

To guarantee the unitarity 
of the TEO in Eq. (\ref{eq:TEONQubit}), 
we deduced the correspondent  constraints 
in appendix \ref{Unit}. 
Writing the TEO coefficients in its polar form $\Lambda_{1}=\left|\Lambda_{1}\right|e^{i \varphi_{1}}$, $\Lambda_{2}=\left|\Lambda_{2}\right|e^{i \varphi_{2}}$ and $\Lambda_{3}=\left|\Lambda_{3}\right|e^{i \varphi_{3}}$, the three constraints we obtained are $\vert\Lambda_{3}\vert=\vert\Lambda_{1}\vert$,  $e^{\mathfrak{Re}(\Lambda_{2})}=1+\vert\Lambda_{1}\vert^{2}$ and $e^{i\mathfrak{Im}(\Lambda_{2})}=-e^{i(\varphi_{1}+\varphi_{3})}\,$. Using the above results, we can rewrite Eq. (\ref{eq:TEONQubit}) as 
\begin{align}
\hat{U}=\frac{1}{\sqrt{1+\left|\Lambda_{1}\right|^{2}}}\begin{bmatrix}
    -e^{i(\varphi_{1}+\varphi_{3})}  & \left|\Lambda_{1}\right| e^{i\varphi_{1}}
   \\
     \left|\Lambda_{1}\right| e^{i\varphi_{3}} & 1
\end{bmatrix}\, ,
\label{eq:TEONQubitfin2}
\end{align}
where we have disregarded the global phase. Notice the dependence on the three real parameters $\left|\Lambda_{1}\right|$, $\varphi_{1}$ and $\varphi_{3}$. 

We have shown that changing the basis of the  Pauli matrices to that of ladder operators yields an invertible coupling matrix for all times in $\mathfrak{su}(2)$ QLSs, allowing us to explicitly calculate the TEO. This is because the bases we defined for the algebras in Table \ref{CommRe}.a are Cartan-Weyl bases, and they naturally align with the root structure of the Lie algebra. In fact, according to Wei and Norman (theorem 5 in Ref. \cite{Wei_1964}), when working with semisimple algebras, which have no non-trivial solvable ideals, this kind of decomposition serves to guarantee an invertible coupling matrix. 

To construct a Cartan-Weyl basis (CWB) for a semisimple Lie algebra, one typically begins by selecting a Cartan subalgebra, which is a maximal abelian subalgebra. It is usual to choose diagonal matrices as its generators. The next step is to construct the associated root spaces, whose elements (raising and lowering operators) are Lie vectors  proportional to themselves when their Lie products with the elements of the Cartan subalgebra are calculated. The  proportionality constants are known as roots, and they come in pairs with opposite sign. In our previous example for $\mathfrak{su}(2)$, $\hat{g}_{1}$ and $\hat{g}_{3}$ can be identified with root vectors from  different root spaces, with roots $+ 1$ and $- 1$, respectively, and $\hat{g}_{2}$ is the only element of the Cartan subalgebra. Of course, the above is not intended to be a formal definition but rather a practical description of a CWB. For an explicit definition we suggest Ref. \cite{Das_2014}, and for a formal definition we suggest Refs.  \cite{Sattinger_1986, Hall_2015}. In Section \ref{symdyn} we will build a generic CWB for the Lie algebra $\mathfrak{su}(N)$, going a bit deeper into this  practical description.

A CWB decomposition provides both theoretical and computational advantages for classical simple algebras, as demonstrated in Refs. \cite{Charzy_ski_2013, Charzy_ski_2015}. From a mathematical perspective, transverse matrices associated with each root subspace 
assume strictly upper or lower triangular form, which are nilpotent. Meanwhile, the matrices corresponding to the Cartan subalgebra are block diagonal. This behaviour is evident in Eqs. (\ref{eq:LieMatrixsuupsil}), for instance. Consequently, all matrix exponentials can be easily computed. When the TEO is expressed as the exponential of the positive (or negative) root, followed by the exponential of the Cartan subalgebra, and then the exponential of the negative (or positive) root, Eq. (\ref{eq:Soldiffeqmatr}) yields a system of block-decoupled nonlinear differential equations  exhibiting a hierarchical structure of coupled Riccati equations in the case of high-order QLSs (or a single Riccati equation for low-order QLSs as in Eq. (\ref{eq:Riccati})). We will observe this structure in the following sections, when working with time-dependent coupled harmonic oscillators in Section \ref{couposc} and $\mathfrak{su}(N)$ in Section \ref{SUNsec}. 

Next, we will evaluate and use our solution of Eq. (\ref{eq:TEONQubitfin2}) in the construction of two important one-qubit quantum gates.


\subsection{$H$ and $T$ single qubit gates}\label{OQgates}

Universal quantum computation can be achieved with the one-qubit Hadamard and $T$ gates, together with the two-qubit CNOT gate \cite{DiVincenzo_1995, Barenco_1995, Bergou_2021}. Here we test and use our solution in Eq. (\ref{eq:TEONQubitfin2}) to demonstrate the construction of the one-qubit mentioned gates.

We consider the time-independent Hamiltonian of a qubit  
driven by a laser (or a microwave for nuclear magnetic resonance) that has detuning $\Delta$ and intensity $\Omega$
\begin{align}
\hat{H} = \frac{\Omega}{2}\hat{\sigma}_{1} + \Delta \frac{\hat{\sigma}_{3}}{2} \,.
\label{eq:rabiHamiltonian}
\end{align}
First, recalling that $\hat{g}_{1}=\frac{1}{2}(\hat{\sigma}_1+i\hat{\sigma}_2)$, $\hat{g}_{3} =\frac{1}{2}(\hat{\sigma}_1-i\hat{\sigma}_2)$ and $\hat{g}_{2}=\frac{1}{2}\hat{\sigma}_{3}$, we can write $\hat{\sigma}_{1}=\hat{g}_{1} + \hat{g}_{3}$ and put the Hamiltonian in the CWB of the previous section, so we can identify $\eta_{1}=\eta_{3}=\Omega/2$ and $\eta_{2}=\Delta$. The qubit state evolves with the differential equation
\begin{align}
    \frac{d}{dt}\expval{\hat{\vb*{\sigma}}} = \vb*{P}\times \expval{\hat{\vb*{\sigma}}},
\end{align}
where $\vb*{P}=(\Omega,0,\Delta)^T$ is a precession vector, and $\expval{\hat{\vb*{\sigma}}} =(\expval{\hat{\sigma}_1},\expval{\hat{\sigma}_2},\expval{\hat{\sigma}_3})^T$ is the Bloch vector~\cite{steck2007quantum}, with $\expval{\hat{\sigma}_i}$ indicating the expected value of operator $\hat{\sigma}_i$.
The $T$ gate is realized by setting the precession vector pointing to the $+z$ direction in the Bloch sphere, and carrying out $1/8$ of a revolution. This is done by setting $\Omega=0$. Further, recalling that the Hamiltonian in Eq.~\eqref{eq:rabiHamiltonian} is time-independent, the TEO can directly be calculated in the unfactorized representation of Eq.~\eqref{eq:unfacTEO}, allowing us to identify $\lambda_j=-i\eta_jt$, with  $j=1,2,3$. Using  the BCH-like equations \eqref{eq:lambda1}-\eqref{eq:lambdamenos}, we get for the $T$ gate $\Lambda_1=\Lambda_3 = 0$ and $\Lambda_2 = -i t\Delta$, where $t$ is the application time of the Hamiltonian, which should be sufficient to carry out one eighth of a revolution, i.e. $t=\pi/(4\Delta)$. Accordingly, using the unitarity constraints we get $\left|\Lambda_{1}\right|=\left|\Lambda_{3}\right|=\mathfrak{Re}(\Lambda_{2})=0$ and $\mathfrak{Im}(\Lambda_{2})=-\pi/4$. Replacing these results in Eq.~\eqref{eq:TEONQubitfin2} shows that
\begin{align}
    \hat{U}_T = \begin{bmatrix}
    e^{-i\pi/4} & 0 \\
    0 & 1
    \end{bmatrix} \,,
\end{align}
or, multiplying by the global phase $e^{i\pi/4}$, we get the canonical representation of the $T$ gate:
\begin{align}
    \hat{U}_T = \begin{bmatrix}
    1 & 0 \\
    0 & e^{i\pi/4}
    \end{bmatrix} \,.
\end{align}

The Hadamard gate $H$ is realized by pointing the precession vector to the direction $(1,0,1)$. \textit{i.e.}, by setting $\Omega = \Delta$. 
This implies that $\lambda_1=\lambda_3=\lambda_2/2:=\lambda$. Replacing in Eqs.~\eqref{eq:lambda1}-\eqref{eq:lambdamenos} we obtain 
\begin{align}
\Lambda_1 = \frac{\sinh(\sqrt{2}\lambda)}{\sqrt{2}\cosh(\sqrt{2}\lambda) - \sinh(\sqrt{2}\lambda)}.
\end{align}
The application time is $t=\pi/(\sqrt{2}\Delta)$, meaning that $\lambda=-i\pi/(2\sqrt{2})$, which implies that $\Lambda_1 = \Lambda_3= -1$ and $\Lambda_2 = \ln{2}- i\pi$. Accordingly, using the unitarity constraints we get $\left|\Lambda_{1}\right|=\left|\Lambda_{3}\right|=1$ and $e^{i\varphi_{1}}=e^{i\varphi_{3}}=-1$. Replacing these results in Eq.~\eqref{eq:TEONQubitfin2} shows that
\begin{align}
     \hat{U}_H = \frac{1}{\sqrt{2}}\begin{bmatrix}
    -1 & -1 \\
    -1 & 1
    \end{bmatrix} \,,
\end{align}
which is the Hadamard matrix up to a global phase of $e^{i\pi}$. The obtainment of the above quantum gates can be understood as a confirmation of our results.

Although recovering the $\hat{\sigma}_3$ gate using Eq. (\ref{eq:TEONQubitfin2}) is straightforward, we note that for the other two individual gates, those proportional to $\hat{\sigma}_1$ and $\hat{\sigma}_2$, it requires considering the limit in which the coupling matrix is singular $\Lambda_{2}\rightarrow\infty$ (see Eq. (\ref{eq:diffeqmatrQLS})). However, this limit can be carefully computed and the mentioned gates can be recovered, as we show in the Appendix~\ref{app:xgateCWB}. Moreover, it is well-known that specific products of the Hadamard and $T$ gates give rise to these gates. Finally, recall that the computation of time-dependent single gates is direct using Eq. (\ref{eq:unfacTEOsol}).


\section{$\texttt{Symdyn}$}\label{symdyn}

Here, we demonstrate the robustness and efficiency of our library in dealing with high-order QLS, by getting key results of the WNM for a quantum system of two coupled time-dependent harmonic oscillators with time-dependent coupling. Furthermore, we use this example to indicate how \verb|symdyn| works. 
We point out that all analytical results of the previous section were computationally recovered using the library, validating it for low-order algebras. The reader can find setting up instructions and examples,  
including those contained in this work, in our repository at \url{https://gitlab.com/VolodyaCO/dynamics-of-sun-systems}.

\begin{figure}[h!]
\includegraphics*[width=6.5cm]{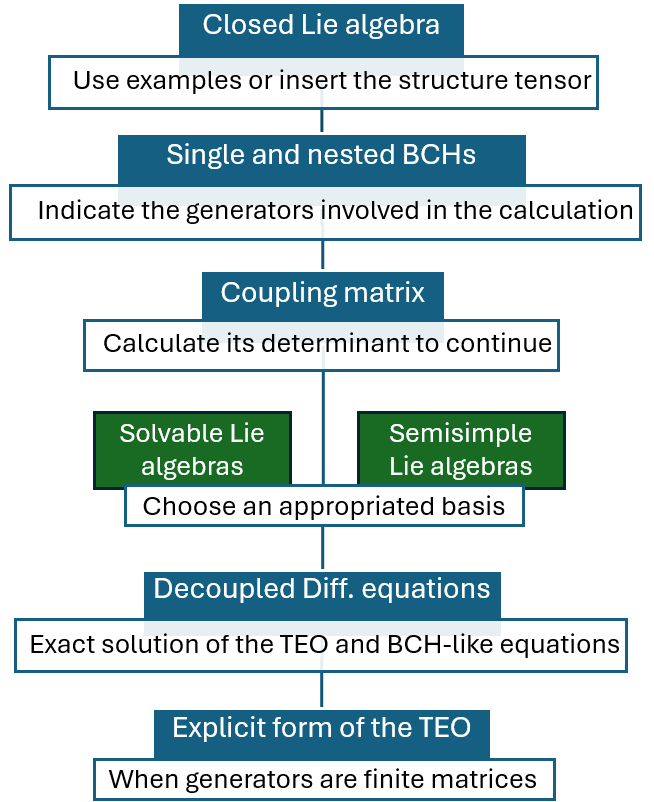}
\caption{Flowchart of the WNM.}
\label{fig:WNMGraph}
\end{figure}

Let us begin by summarizing the key results of the WNM, corresponding to the shadow boxes of the flowchart in Fig. \ref{fig:WNMGraph}. The transparent ones indicate the interaction between the user and \verb|symdyn|. 
%
As it can be seen, sometimes the user must specify the objects involved in the calculation to obtain the desired information, and other times only needs to request the information. 
Be aware that the last two steps of the flowchart are only possible depending on the chosen basis of the Lie Algebra with which you are working.
\verb|Symdyn|'s main class is called \verb|Algebra|, which can be imported in Python as follows
\begin{python}
from symdyn.algebra import Algebra
\end{python}
%
To instantiate an \verb|Algebra|, a structure tensor object must be passed.
The user will have two ways to define the structure tensor. In case the generators are finite matrices, the user can input them one by one, and ask \verb|symdyn| to compute the corresponding commutators to build $\gamma$ for a specific ordering of the basis. We will explore this option for $\textit{SU}(3)$ in Sec. \ref{SUNsec}. The other option is to input the structure tensor $\gamma$ directly, as we show in the following example. 

\subsection{Two time-dependent coupled harmonic oscillators}\label{couposc}

The quantum harmonic oscillator is one of the most important systems in quantum physics. As already mentioned, its time-dependent version allows the study of squeezing, which consists in reducing the uncertainty of one of the quadratures of the oscillator below the value for coherent states (for the other quadrature, the uncertainty increases accordingly) by using the non-adiabatic variation of the oscillator parameters (mass and frequency). This effect is relevant in many fields of physics \cite{WALLS-1983, LOUDON-1987, WU-1987, TEICH-1989, GRISHCHUK-1990, Lo-ST-1991, PERINA-1991, GRISHCHUK-1993,  ALBRECHT-1994,  HU-1994, DODONOV-2002, DODONOV-2003, EINHORN-2003,  DODONOV-2005, FUJII-2011, LAHTEENMAKI-2013, FELICETTI-2014,  SCHNABEL-2017, Raffa2019, Leroux-2010, Hosten-2016,  Bao2020}, and useful to enhancing the sensitivity of several systems \cite{VAHLBRUCH-2007, GIOVANNETTI-2011}, including  telecommunications \cite{SLAVIK-2010, Fedorov2016, Pogorzalek2019} and the detection rate at LIGO \cite{Barsotti_2018, Tse_2019, Jia_2024}.

Our first high-order QLS consist of two time-dependent harmonic oscillators coupled with a time-dependent coupling.  The most general case involves a total of 15 generators and all time-dependent physical parameters  \cite{Wolf_1988}. 
A simplified version of this problem was recently solved using the WNM together with a numerical approach by I. Ramos-Prieto  \textit{et. al.} \cite{Ramos_2021}, in the context of effective approaches to the dynamical Casimir effect. The system consists of a one-dimensional cavity with a moving mirror in one boundary and two coupled modes of the electromagnetic field inside \cite{Law_1994}. They 
reduced the number of generators to 11 by ignoring terms proportional to single ladder operators of each mode. In our development, we 
hold this reduction 
for the simplicity of the expressions. The effective Hamiltonian of the system can be written as 
\begin{equation}
\hat{H}(t)=\sum_{l=1}^{11}\eta_{l}(t)\hat{g}_{l} \, ,
\label{eq:HamCoup}
\end{equation}
where the coefficients are functions of the physical parameters of the system and the generators are bilinear products of ladder operators of each mode (see Appendix \ref{appD}),  forming a CWB for the system. 

Our first step is to calculate the Structure Tensor,
which can be deduced from the structure table of the system,  Table  \ref{CommRecoupled} in Appendix \ref{appD}. Notice that some Lie products are proportional to a linear combination of two or more generators, unlike in our previous examples where they were proportional to a single generator. To encode $\gamma$ in \verb|symdyn|, one must initially define the algebraic order $L$ to create a zero-filled $L\times L\times L$ tensor. For our current system $L=11$, and the code to generate it is
\begin{python}
import numpy as np
gamma = np.zeros((11, 11, 11))
\end{python}
We then proceed to indicate the nonzero entries of each transverse matrix. For instance, the top transverse matrix, corresponding to the first row of the structure matrix in Table  \ref{CommRecoupled}, is encoded as
\begin{python}
gamma[0, 4, 2] = -1 
gamma[0, 5, 0] = -2 
gamma[0, 7, 5] = -4  
gamma[0, 7, 10] = -2  
gamma[0, 9, 3] = -2 
\end{python}
where it must be noticed that Python counts elements of collections from zero. In this way, the first line in the above code 
indicates that the entry correspondent to the 
commutator between $\hat{g}_{1}$ and $\hat{g}_{5}$ (first two indexes) is proportional to $\hat{g}_{3}$ (third index), with coefficient of proportionality $-1$. In case a commutator is proportional to more than one generator, new lines (with the same first two indexes) must be written  specifying each generator with its proportionality coefficient. For example, in the above code the third and fourth lines indicate that the commutator between $\hat{g}_{1}$ and $\hat{g}_{8}$ is given by the linear combination of the generators $\hat{g}_{6}$ and $\hat{g}_{11}$, with coefficients of proportionality $-4$ and $-2$, respectively. Importantly, we just need to fill the terms above (or below) the diagonal of the structure matrix. \verb|Symdyn| automatically calculates the missing terms using the anti-symmetry of the commutator. 
It is worth noting that we can incorporate symbolic entries  in $\gamma$. As an example, for the low-order algebras of the previous section, using Table  \ref{CommRe}.a we can code $\gamma$ including $\epsilon$ and $\upsilon$ as 
\begin{python}
import sympy
epsilon = sympy.Symbol("\\epsilon")
upsilon = sympy.Symbol("\\upsilon")
gamma[0, 1, 0] = -upsilon
gamma[0, 2, 1] = -2*epsilon
gamma[1, 2, 2] = -upsilon
\end{python}
Once $\gamma$ has been introduced, we can initialize the \verb|Algebra| class as
\begin{python}
algebra = Algebra(gamma)
\end{python}

To continue benchmarking the library, a single similarity transformation, for instance $e^{\Lambda_{2}\hat{g}_{2}}\hat{g}_{4}\, e^{-\Lambda_{2}\hat{g}_{2}}$, can be calculated using the code
\begin{python}
algebra.get_similarity_transform(2, 4)
\end{python}
from which \verb|symdyn| will return a tuple with both an expression in \LaTeX \,and the resulting Lie vector. In this case we get
\begin{align}
-2\Lambda_{2}\hat{g}_3 + \hat{g}_4,
\end{align}
as expected. Other complex calculations, as nested similarity transformations of Lie vectors
\begin{align}
(e^{\Lambda_{k}\text{ad}\hat{g}_{k}})(e^{\Lambda_{j}\text{ad}\hat{g}_{j}})\cdots(e^{\Lambda_{i}\text{ad}\hat{g}_{i}})\sum_{l=1}^{L}\chi_{l}\hat{g}_{l} \, , 
\label{eq:notsameas}
\end{align}
where $i,j,k\in\{1,2,\ldots,L\}$, can also be calculated with the library\footnote{Note that this type of calculation allows to obtain changes of basis, the time evolution of any Lie vector within the Heisenberg picture and, more generally, for suitable well-behaved functions of the generators $f(\hat{g}_{j})$, we have that $(e^{\Lambda_{i}\text{ad}\hat{g}_{i}}) f(\hat{g}_{j})=f((e^{\Lambda_{i}\text{ad}\hat{g}_{i}}) \hat{g}_{j})$ \cite{Barnett-1997}.}. See the GitLab repository for further details on the usage of this routine, available under \verb|algebra.get_nested_similarity_transform|.
%

The next key result of the WNM are the BCH matrices, which are obtained using the code
\begin{python}
b_matrices = algebra.b_matrices
\end{python}
displaying all possible similarity transformations between the generators. Due to the size of these matrices, we prefer not to show them explicitly here, although the reader can find them in our repository. The next key result of the WNM is the coupling matrix $\vb*{\xi}$, which is obtained using the code 
\begin{python}
algebra.xi_matrix
\end{python}
Although we will not explicitly show this matrix here for the same reason than for the BCH matrices, we calculated it together with the coupled equations resulting from Eq. (\ref{eq:diffeqmatr}) in our library, which are obtained using the code
\begin{python}
algebra.get_coupled_differential_equations()
\end{python}
These coupled equations agree with those in Ref. \cite{Ramos_2021}, so we validate our numerical implementation for high orders. 

We can go further and calculate the determinant of the coupling matrix to find out if the solution is global. This is done with the code
\begin{python}
sympy.det(algebra.xi_matrix)
\end{python}
For our current system we obtain $\det\vert\boldsymbol{\xi}\vert = \exp{-3(\Lambda_{6}+\Lambda_{7})}\,$, which verifies to be non-singular for all times except in the limit $(\Lambda_{6}+\Lambda_{7}) \rightarrow\infty$.   Accordingly, we can continue with the WNM in \verb|symdyn| to obtain the system's block-decoupled differential equations (\ref{eq:Soldiffeqmatr}). This is done using the code 
\begin{python}
algebra.get_decoupled_differential_equations()
\end{python}
which, for our current example, returns
\begin{small}
\begin{align}
\label{eq:eta1findeCoup}
i \dot{\Lambda}_{1} =& 4 \Lambda_{1}^{2} \eta_{8} + 2 \Lambda_{1} \Lambda_{3} \eta_{10} + 2 \Lambda_{1} \eta_{6} + \Lambda_{3}^{2} \eta_{9} + \Lambda_{3} \eta_{4} + \eta_{1} \, ,\\
\label{eq:eta2findeCoup}
i \dot{\Lambda}_{2} =& 4 \Lambda_{2}^{2} \eta_{9} + 2 \Lambda_{2} \Lambda_{3} \eta_{10} + 2 \Lambda_{2} \eta_{7} + \Lambda_{3}^{2} \eta_{8} + \Lambda_{3} \eta_{5}  + \eta_{2} \,,\\
\label{eq:eta3findeCoup}
i \dot{\Lambda}_{3} =& \Lambda_{3}^{2} \eta_{10} + 4\Lambda_{3} (\Lambda_{1} \eta_{8} + \Lambda_{2}\eta_{9})  + \Lambda_{3}(\eta_{6} + \eta_{7}) + \nonumber\\[2pt]
& + 2\Lambda_{2}(2\Lambda_{1}  \eta_{10}  + \eta_{4})  + 2\Lambda_{1}\eta_{5} + \eta_{3} \,,
\end{align}
\begin{align}
\label{eq:eta4findeCoup}
i \dot{\Lambda}_{4} =& - \Lambda_{4}^{2} (2\Lambda_{3} \eta_{8} + 2\Lambda_{2} \eta_{10} + \eta_{5}) - 4 \Lambda_{4} (\Lambda_{2} \eta_{9}  -  \Lambda_{1} \eta_{8})  + \nonumber\\[2pt]
&  + \Lambda_{4} (\eta_{6} - \eta_{7}) + 2 \Lambda_{3} \eta_{9}  + 2 \Lambda_{1} \eta_{10} + \eta_{4} \,,\\
\label{eq:eta5findeCoup}
i \dot{\Lambda}_{5} =&  2 \Lambda_{4} \Lambda_{5} (2\Lambda_{3} \eta_{8}  +  2\Lambda_{2} \eta_{10} + \eta_{5}) + 4 \Lambda_{5} (\Lambda_{2}\eta_{9} - \Lambda_{1} \eta_{8}) + \nonumber\\[2pt]
&  +  \Lambda_{5} (\eta_{7} - \eta_{6}) + 2 \Lambda_{3} \eta_{8} + 2 \Lambda_{2} \eta_{10}  + \eta_{5} \,,\\
\label{eq:eta6findeCoup}
i \dot{\Lambda}_{6} =&  -\Lambda_{4} (2 \Lambda_{3} \eta_{8} + 2 \Lambda_{2}\eta_{10} + \eta_{5}) +  4 \Lambda_{1} \eta_{8}  + \Lambda_{3} \eta_{10} + 
\nonumber\\[2pt]
&   +  \eta_{6}\,,\\
\label{eq:eta7findeCoup}
i \dot{\Lambda}_{7} =&  \Lambda_{4} (2 \Lambda_{3} \eta_{8} + 2 \Lambda_{2} \eta_{10} +  \eta_{5}) + 4 \Lambda_{2} \eta_{9}  + \Lambda_{3} \eta_{10} + \eta_{7}
,\\
\label{eq:eta8findeCoup}
i \dot{\Lambda}_{8} =&  \Lambda_{5}^{2}(\Lambda_{4}^{2} \eta_{8} + \Lambda_{4}\eta_{10} + \eta_{9})e^{2 \Lambda_{6}} +  \Lambda_{5} (2 \Lambda_{4} \eta_{8} + \eta_{10})e^{2 \Lambda_{6}} + \nonumber\\[2pt]
&   + \eta_{8} e^{2 \Lambda_{6}} \,,\\
\label{eq:eta9findeCoup}
i \dot{\Lambda}_{9} =& \left(\Lambda_{4}^{2} \eta_{8} + \Lambda_{4} \eta_{10} + \eta_{9}\right) e^{2 \Lambda_{7}} 
\,,\\
\label{eq:eta10findeCoup}
i \dot{\Lambda}_{10} =&  2 \Lambda_{5}( \Lambda_{4}^{2}  \eta_{8} + \Lambda_{4} \eta_{10} +  \eta_{9})e^{\Lambda_{6}+\Lambda_{7}} + 2 \Lambda_{4} \eta_{8}e^{\Lambda_{6}+\Lambda_{7}} + 
\nonumber\\[2pt]
&   + \eta_{10} e^{\Lambda_{6}+\Lambda_{7}} 
\,,\\
\label{eq:eta11findeCoup}
i \dot{\Lambda}_{11} =& 2 \Lambda_{1} \eta_{8} + 2 \Lambda_{2} \eta_{9} + \Lambda_{3} \eta_{10}  + \eta_{11} 
\, . 
\end{align}
\end{small}
Note that the first three equations form a block of coupled Riccati equations, which once solved, allows solving the fourth equation, which is also a Riccati equation. Finally, the rest can be solved by quadrature. This hierarchy of Riccati equations is expected, once the basis for the Lie algebra is a CWB \cite{Charzy_ski_2013, Charzy_ski_2015}.
Also note that the determinant of the coupling matrix is given by the exponential of a linear combination of coefficients $\Lambda_6$ and $\Lambda_7$, being the ones related with the generators of the Cartan subalgebra. 
To the authors' knowledge, this is the first time that the above equations have been calculated. 

Finally, as mentioned before, changing the order for the exponential generators in the TEO involves recalculating $\gamma$. For instance, let us consider the new configuration for the TEO of our current example
\begin{align}
\hat{U}=& 
e^{\Lambda_{1}\hat{g}_{5}}e^{\Lambda_{2}\hat{g}_{1}}e^{\Lambda_{3}\hat{g}_{2}}e^{\Lambda_{4}\hat{g}_{7}}e^{\Lambda_{5}\hat{g}_{11}}e^{\Lambda_{6}\hat{g}_{3}}e^{\Lambda_{7}\hat{g}_{4}}e^{\Lambda_{8}\hat{g}_{9}} \nonumber\\[2pt]
&e^{\Lambda_{9}\hat{g}_{6}}e^{\Lambda_{10}\hat{g}_{8}}e^{\Lambda_{11}\hat{g}_{10}}\, .
\label{eq:newteo}
\end{align}
To calculate the new  $\gamma$ it must be introduced the code
\begin{python}
new_ord = (5, 1, 2, 7, 11, 3, 4, 9, 6, 8, 10)
new_alg = algebra.change_generators_order(new_ord)
\end{python}
This code swaps the rows, columns and layers of the initial $\gamma$ accordingly, and thus a new \verb|Algebra| object with the new $\gamma$ is created. After this, the user can obtain key results of the WNM from the new $\gamma$ in the same way we did before. 

In the next section, we specialize our library in the Lie group $\textit{SU}(N)$ and show an example of the direct introduction of finite matrix generators in the library to configure the Structure Tensor of the system. 


\section{$\textit{SU}(N)$}\label{SUNsec}

Without a doubt, $\textit{SU}(N)$ is one of the most important groups in physics, playing a special role in quantum mechanics, quantum field theory, and the standard model of particle physics \cite{Pfeifer_2003}. In particular, $\textit{SU}(2)$ and $\textit{SU}(3)$ describe key aspects of particle interactions as isospin symmetry \cite{Sakurai-Book-2014} and the strong force among quarks and gluons \cite{Skands_2013}, respectively. 
Larger groups like $\textit{SU}(5)$ have been used to extend these symmetries in attempts to unify all fundamental interactions in grand unified theories  \cite{Georgi_1974}. 

In quantum computation, the role of $\textit{SU}(N)$ is foundational, as a mathematical structure for describing quantum gates and transformations, and providing tools for error correction and algorithms optimization \cite{Nielsen_2012, Bergou_2021, Wiersema_2024}. For qubit-based quantum computing, the special case where the group dimension is a power of two ($N = 2^n$) is considered, with $n$ indicating the number of qubits in the system. Indeed, and as we showed in Section \ref{OQgates}, the case $n=1$ is the usual algebra to describe a time-dependent qubit.  On the other hand, $\textit{SU}(3)$ has also been used to describe quantum computing but with qutrits, which allow more efficient encoding of information and computations \cite{Caves_2000, Klimov_2003, Gokhale_2019}. Higher-order computing with qu$d$its, described by $SU(d)$, is also possible  \cite{Wang_2020}.

In this section, we present a generic CWB for $\mathfrak{su}(N)$ and show how it can be leveraged in \verb|symdyn| using $\mathfrak{su}(2)$ as an example (Section~\ref{GCWB}). 
We then use the library to obtain some key results of the method for $\textit{SU}(3)$ and $\textit{SU}(4)$, including the TEO in its explicit matrix form. We use $\textit{SU}(3)$ to exemplify the explicit introduction of finite matrix generators into the library (Section~\ref{sec:su3}). We finish by computing the CNOT gate, completing the gates needed for universal quantum computation (Section~\ref{sec:su4}).

\subsection{Generic CWB for $\mathfrak{su}(N)$}\label{GCWB}

Here, we define a generic CWB for $\mathfrak{su}(N)$, allowing us to apply the WNM in \verb|symdyn| for an arbitrary $N$. Recall that this algebra comprises $L=N^{2}-1$ generators, which are $N\times N$ traceless matrices. To establish a CWB, we start by constructing a Cartan subalgebra $\mathfrak{H}$, which is composed 
of the maximum number of diagonalizable $N\times N$ matrices, forming a maximal commutative (Abelian) algebra \cite{Charzy_ski_2013, Charzy_ski_2015}.
By the traceless condition, the number of generators in $\mathfrak{H}$ is $N-1$, defined as the rank of the algebra.
Let us identify these generators by $\hat{h}_{j}$, so that $j=1,2,\ldots ,N-1$. One simple option, is to define them as diagonal matrices with two nonzero entries forming blocks of $1$ and $-1$, namely,
\begin{align}
\hat{h}_{j} \equiv \text{diag}(0, \ldots, 0, \underbrace{1}_{j}, \underbrace{-1}_{j+1}, 0, \ldots, 0) \,.   
\label{eq:Cartan}
\end{align}
%
On the other hand, the generators of the root spaces correspond to the off-diagonal $N\times N$ matrices $\hat{r}_{nk}$, with all zero elements except a single $1$ in entry $(n, k)$, representing raising ($n < k$) and lowering ($n > k$) operators. The number of generators of each root space is given by half the difference between the algebraic order and its rank, namely, $\frac{N(N-1)}{2}$. Finally, we choose the ordering of the generators to be the following:
\begin{align}
\hat{g}_{l} = 
    \begin{cases}
   \hat{r}_{nk}, & \text{for} \,l=f_{1}(n,k), \,\, \mbox{with} \,\, n<k
   		\\
        \hat{h}_{j}, & \text{for } 
        l =\frac{N(N-1)}{2}+j, \\
        	\hat{r}_{nk}, & \text{for} \,    				l=\frac{N^{2}+N-2}{2}+f_{2}(n,k), \,\, 
        	\mbox{with} \,\,
   		n>k,
    \end{cases} 
\label{eq:gensunordering}
\end{align}
where $f_{1}(n,k)=k+N(n-1)-\frac{n}{2}(n+1)$ and $f_{2}(n,k)=k + \frac{1}{2}(n-1)(n-2)$, with $n,k=1,2,\ldots,N$. 
Notice that we generically ordered the basis such that positive roots generators are first, second are the Cartan subalgebra generators, and the negative roots generators are third. 

The above results provide a way to systematically construct this CWB for any $N$ of $\mathfrak{su}(N)$. In \verb|symdyn|, these generators can be built as follows
\begin{python}
from symdyn.sun.utils import get_sun_generators
generators = get_sun_generators(N)
\end{python}
with \verb|N| specified by the user. Then, the Structure Tensor of the system is built as
\begin{python}
from symdyn.utils import get_gamma
gamma = get_gamma(generators)
\end{python}
%
and the TEO in its explicit form 
can be obtained with the code 
\begin{python}
from symdyn import get_TEO
get_TEO(generators)
\end{python}
%

As an example, but also for the sake of completeness, let us obtain from \verb|symdyn| the generators and the decoupled equations for $\mathfrak{su}(2)$ with the above defined CWB. First, we use the code
\begin{python}
algebra.get_sun_generators(2)
\end{python}
to load the basis. The generators can be printed, and they are
\begin{align}
\hat{g}_1 =& 
\begin{bmatrix}
    0 & 1 
   \\
     0 & 0
\end{bmatrix}  , \,\,\, 
\hat{g}_2 =  \begin{bmatrix}
  1 & 0 
   \\
     0 & -1
\end{bmatrix} , \,\,\,
\hat{g}_3 =  \begin{bmatrix}
    0 & 0 
   \\
     1 & 0
\end{bmatrix}  \,.
\label{eq:su2CWB}
\end{align}
Using this basis, it can be built the Structure Tensor, and from it  the coupling matrix, which determinant is $\det\vert\boldsymbol{\xi}\vert = \exp{-2\Lambda_{2}}\,$, and consequently we can obtain the decoupled equations
\begin{align}
\label{eq:su(2)CWB}
&i\dot{\Lambda}_{1}=-\Lambda_{1}^{2}\eta_{3}+ 2 \eta_{2}\Lambda_{1} +\eta_{1} \, , \\
&i\dot{\Lambda}_{2}=- \Lambda_{1}\eta_{3}+ \eta_{2} \, , \\
&i\dot{\Lambda}_{3}=\eta_{3} e^{2\Lambda_{2}} \,.
\end{align}
The first equation is a Riccati equation, which once solved, allows solving the rest by quadrature. This is the expected structure of the differential equations once the basis for the Lie algebra is a CWB. Note that the determinant of the coupling matrix only depends on the generator of the Cartan subalgebra. Finally, the explicit TEO is given by
\begin{align}
\hat{U}=e^{-\Lambda_{2}}\begin{bmatrix}
    e^{2\Lambda_{2}}+ \Lambda_{1}\Lambda_{3} & \Lambda_{1} 
   \\
     \Lambda_{3} & 1
\end{bmatrix}\, .
\label{eq:TEONQubitCWB}
\end{align}
Note that the basis in Eq. (\ref{eq:su2CWB}) differs from that in Table \ref{CommRe} for $\mathfrak{su}(2)$ only by a constant in the generator of the Cartan subalgebra $\hat{g}_2$. Consequently, the determinant of the coupling matrices and the explicit TEOs are also different in these bases.  



\subsection{$\mathfrak{su}(3)$}\label{sec:su3}

As mentioned before, $\mathfrak{su}(3)$ is a relevant algebra to describe many important physical systems. Further, it is usual to use the Gell-Mann matrices as its basis. However, by its cyclic structure (this basis can be considered as a generalization of the Pauli basis), we will need to change it to a CWB to get a factorized version of the correspondent TEO. 

The Gell-Mann matrices are:
\begin{align}
\underline{\hat{g}}_{1}=&
 \begin{bmatrix}
    0 & 1 & 0
 \\     
    1 & 0 &  0 
   \\
        0 & 0 & 0
\end{bmatrix},  \,\, 
\underline{\hat{g}}_{2}=\begin{bmatrix}
      0 & -i & 0
 \\     
    i & 0 &  0 
   \\
        0 & 0 & 0
\end{bmatrix} , \,\,
\underline{\hat{g}}_{3}=
\begin{bmatrix}
      1 & 0 & 0
 \\     
    0 & -1 &  0 
   \\
        0 & 0 & 0
\end{bmatrix} , 
\nonumber \\
\underline{\hat{g}}_{4}=&
 \begin{bmatrix}
    0 & 0 & 1
 \\     
   0 & 0 &  0 
   \\
        1 & 0 & 0
\end{bmatrix},  \,\, 
\underline{\hat{g}}_{5}=\begin{bmatrix}
        0 & 0 & -i
 \\     
   0 & 0 &  0 
   \\
        i & 0 & 0
\end{bmatrix} , \,\,
\underline{\hat{g}}_{6}=
\begin{bmatrix}
    0 & 0  &  0
 \\     
   0 & 0 & 1 \\
        0 & 1 & 0
\end{bmatrix} , \nonumber \\
&\underline{\hat{g}}_{7}=
 \begin{bmatrix}
       0 & 0  &  0
 \\     
   0 & 0 & -i \\
        0 & i & 0
\end{bmatrix},  \,\,\, 
\underline{\hat{g}}_{8}=\frac{1}{\sqrt{3}}\begin{bmatrix}
    1 & 0 & 0
 \\     
   0 & 1 & 0 
   \\
     0 & 0 & -2
\end{bmatrix} , 
\label{eq:Gell}
\end{align}
satisfying the Lie products
\begin{equation}
[\underline{\hat{g}}_{j},\underline{\hat{g}}_{k}]=2i\sum_{l=1}^{8}f_{jkl}
\underline{\hat{g}}_{l} \, , 
\label{eq:adop}
\end{equation}
with $\gamma_{jkl}=2if_{jkl}$ the structure constants. Of the latter, the only ones that have a value different from zero are
\begin{align}
f_{123} &= 1\, ,\,\,f_{458} = f_{678} = \frac{\sqrt{3}}{2}\, ,\\
f_{147} &= f_{165} = f_{246} = f_{257} = f_{345} = f_{376} = \frac{1}{2}\, ,  
\label{eq:f_ijk}
\end{align}
being completely antisymmetric in all three indices. 
Accordingly, the Hamiltonian can be written as
\begin{equation}
\hat{H}(t)=\sum_{l=1}^{8}\varsigma_{l}(t)\underline{\hat{g}}_{l} \, ,
\label{eq:Hamsu3}
\end{equation}
and the correspondent TEO as
\begin{align}
\hat{U}(t) = \prod_{l=1}^{8} e^{ \Theta_{l}(t) \underline{\hat{g}}_{l}}\, .
\label{eq:TEOsu3paili}
\end{align}
%


To introduce the generators in Eqs. (\ref{eq:Gell}) in the library, \verb|L| zero-filled matrices of dimension \verb|N| are be created with the code 
%
\begin{python}
generators = np.zeros((L, N, N))
\end{python}
In our case, \verb|L|$=8$ and \verb|N|$=3$. Then, the entries that are different from zero in each generator must be informed. For instance, generators $\hat{g}_{1}$ and $\hat{g}_{8}$ in Eqs. (\ref{eq:Gell}) are introduced as
\begin{python}
# First generator:
generators[0, 0, 1] = 1 
generators[0, 1, 0] = 1
# Eighth generator:
generators[7, 0, 0] = 1 / np.sqrt(3)
generators[7, 1, 1] = 1 / np.sqrt(3)
generators[7, 2, 2] = -2 / np.sqrt(3)
\end{python}
Once all the generators has been introduced, \verb|symdyn| automatically calculates their Lie products\footnote{In case the inserted matrices do not close a Lie algebra, the user will be notified by the library.}, and with them the correspondent Structure Tensor. When working with higher-order Lie algebras with generators satisfying cyclic Lie products, such as Gell-Mann matrices, this property can be used to automate the construction of the basis, as we show in our repository.

We can proceed with calculating the system's coupling matrix and its determinant. Similar to  the Pauli basis for $\mathfrak{su}(2)$, the Gell-Mann basis produces a coupling matrix with periodic singularities that prevents the solution to be global. Owing to the size of this matrix, we have chosen not to display it explicitly here; however, it is available in our repository. As a consequence, a global factorized version of the TEO cannot be obtained using this basis. Nevertheless, using a CWB the situation changes. For example, in one of the pioneering applications of the WNM for this algebra \cite{Dattoli_1987}, the authors considered the group $\textit{SU}(2)\times\textit{SU}(2)\times \textit{SU}(2)$ instead of $\textit{SU}(3)$ directly, once the latter can be obtained from the former through a transformation, and used a CWB for the underlying algebra of each $\textit{SU}(2)$ subgroup, effectively simplifying the problem.

Taking $N=3$ in Eq. (\ref{eq:gensunordering}), we obtain the following CWB for $\mathfrak{su}(3)$:
\begin{align}
\hat{g}_{1}=&
 \begin{bmatrix}
   0 & 1 & 0 \\
	0 & 0 & 0 \\
	 0 & 0 & 0
\end{bmatrix},  \,\, 
\hat{g}_{2}=\begin{bmatrix}
     0 & 0 & 1 \\
	0 & 0 & 0 \\
	 0 & 0 & 0
\end{bmatrix} , \,\,
\hat{g}_{3}=
\begin{bmatrix}
     0 & 0 & 0 \\
	0 & 0 & 1 \\
	 0 & 0 & 0
\end{bmatrix} , \nonumber \\
&\hat{g}_{4}=
 \begin{bmatrix}
   1 & 0 & 0 \\
0 & - 1 & 0 \\
0 & 0 & 0
\end{bmatrix},  \,\,\, 
\hat{g}_{5}=\begin{bmatrix}
    0 & 0 & 0 \\
0 & 1 & 0 \\
0 & 0 & -1
\end{bmatrix} ,
\nonumber \\
\hat{g}_{6}=&
 \begin{bmatrix}
  0 & 0 & 0 \\
	1 & 0 & 0 \\
	 0 & 0 & 0
\end{bmatrix},  \,\, 
\hat{g}_{7}=\begin{bmatrix}
 0 & 0 & 0 \\
	0 & 0 & 0 \\
	 1 & 0 & 0
\end{bmatrix} , \,\,
\hat{g}_{8}=
\begin{bmatrix}
  0 & 0 & 0 \\
	0 & 0 & 0 \\
	 0 & 1 & 0
\end{bmatrix} ,  
\label{eq:CWBsu3}
\end{align}
which can be loaded in \verb|symdyn| with 
\verb|generators = get_sun_generators(N)|. 
The TEO now reads
\begin{align}
\hat{U}(t) = \prod_{l=1}^{8} e^{ \Lambda_{l}(t)\hat{g}_{l}} \, .
\label{eq:TEOSU3c}
\end{align}
We can continue to ask \verb|symdyn| for the coupling matrix of the system (not shown here) and its determinant, the latter resulting in $\det\vert\boldsymbol{\xi}\vert = \exp{-2(\Lambda_{4}+\Lambda_{5})}\,$, involving the coefficients of the generators belonging to the Cartan subalgebra. Accordingly, we can ask \verb|symdyn| for the system of block-decoupled differential equations, which returns
\begin{small}
\begin{align}
\label{eq:RicEqSU3-1}
i \dot{\Lambda}_{1} =& -\Lambda_1^2 \eta_6 - \Lambda_1 \Lambda_2 \eta_7 + \Lambda_1 (2\eta_4 - \eta_5) - \Lambda_2 \eta_8 + \eta_1 \, ,\\
\label{eq:RicEqSU3-2}
i \dot{\Lambda}_{2} =& - \Lambda_2^2 \eta_7 -\Lambda_1 \Lambda_2 \eta_6   + \Lambda_2 (\eta_4 + \eta_5) - \Lambda_1 \eta_3 + \eta_2 \,,\\
\label{eq:RicEqSU3-3}
i \dot{\Lambda}_{3} =& -\Lambda_3^2 (\eta_8 - \Lambda_1 \eta_7) + \Lambda_3 (\Lambda_1 \eta_6 - \Lambda_2 \eta_7) +  \nonumber\\[2pt]
& + \Lambda_3 (2 \eta_5  - \eta_4)+ \Lambda_2 \eta_6 + \eta_3 \,,\\
\label{eq:RicEqSU3-4}
i \dot{\Lambda}_{4} =& -\Lambda_1 \eta_6 - \Lambda_2 \eta_7 + \eta_4 \,,\\
\label{eq:RicEqSU3-5}
i \dot{\Lambda}_{5} =&  -\Lambda_1 \Lambda_3 \eta_7 - \Lambda_2 \eta_7 - \Lambda_3 \eta_8 + \eta_5  \,,\\
\label{eq:RicEqSU3-6}
i \dot{\Lambda}_{6} =& (\eta_6 - \Lambda_3 \eta_7)e^{(2 \Lambda_4- \Lambda_5)} \,,\\
\label{eq:RicEqSU3-7}
i \dot{\Lambda}_{7} =& \Lambda_6(\Lambda_1  \eta_7 +  \eta_8) e^{(- \Lambda_4 + 2 \Lambda_5)} + 
\eta_7 e^{( \Lambda_4 + \Lambda_5)} 
,\\
\label{eq:RicEqSU3-8}
i \dot{\Lambda}_{8} =&  (\Lambda_1 \eta_7  + \eta_8) e^{(- \Lambda_4 + 2 \Lambda_5)}
\, . 
\end{align}
\end{small}Note that the first two equations (the same number that the rank of this algebra) form a block of coupled Riccati equations, which once solved, allows solving the third equation, which is also a Riccati equation. Then, the rest can be solved by quadrature. As mentioned, this hierarchy of Riccati equations is expected. 
Finally, the explicit TEO is obtained
\begin{small}
\begin{align}
\hat{U}=e^{-\Lambda_{5}}\begin{bmatrix}
   \Lambda_{1}\Lambda_{6}\varrho_{1} + \Lambda_{7}\varrho_{2} + e^{\Lambda_{4}\Lambda_{5}} & \Lambda_{1}\varrho_{1} + \Lambda_{8}\varrho_{2} & \varrho_{2} \\
	\Lambda_{3}\Lambda_{7}+\Lambda_{6}\varrho_{1} & \Lambda_{3}\Lambda_{8}+\varrho_{1} & \Lambda_{3} \\
	 \Lambda_{7} & \Lambda_{8} & 1
\end{bmatrix}\, ,
\label{eq:TEONsu3CWB}
\end{align}
\end{small}where $\varrho_{1}=e^{-\Lambda_{4}+2\Lambda_{5}}$ and $\varrho_{2}=\Lambda_{1}\Lambda_{3}+\Lambda_{2}$.

\subsection{$\mathfrak{su}(4)$ and beyond}\label{sec:su4}

Here, we concentrate in the calculation of the  CNOT gate. Other key results of the WNM for the Lie algebra $\mathfrak{su}(4)$, as the explicit TEO or the system of block-decoupled differential equations, can be found in Appendix \ref{sec:su4B}.

The CNOT gate is given by the $4\times 4$ matrix 
\begin{align}
    \hat{U}_{CNOT} = \begin{bmatrix}
   1 & 0 & 0 & 0 \\
	0 & 1 & 0 & 0 \\
	 0 & 0 & 0 & 1 \\
	  0 & 0 & 1 & 0
\end{bmatrix} \,. 
\label{eq:CNOTM}
\end{align}
We focus on finding the $\Lambda_i$ coefficients that yield the above gate.
The calculations are merely algebraic, and can be found in Appendix \ref{sec:su4B}. 
There, we introduce a global phase $\phi_n$ to work directly with the TEO coefficients without applying the unitary condition. This phase is shown to be $\phi_n=\frac{(2n+1)\pi}{4}$ with  $n\in\mathbb{Z}$, and the TEO coefficients to satisfy $\Lambda_{6}= \Lambda_{15}=e^{\Lambda_{9}+i\phi_n}$, $\Lambda_{7}=i\phi_n$, $\Lambda_{8}=i\pi(\frac{1}{2}-n)$, $\Lambda_{11}= \Lambda_{13}e^{(-i\phi_n-\Lambda_9)}$, $\Lambda_{12}= \Lambda_{14}e^{(-i\phi_n-\Lambda_9)}$, while $\Lambda_{1}=\Lambda_{2}=\Lambda_{3}=\Lambda_{4}=\Lambda_{5}=\Lambda_{10}=0$, and $\Lambda_{13}$ and $\Lambda_{14}$ are free complex parameters. In the limit $\Lambda_{9}\rightarrow\infty\,$, the TEO matches the CNOT gate up to the global phase $\phi_n$.  

The construction of the CNOT gate, as well as the $H$ and $T$ gates in the case of one qubit (section~\ref{OQgates}) shows how the WNM coefficients can be systematically obtained in an example where we have studied a set of universal quantum gates for quantum computing.
We anticipate that, due to the complexity of the block-decoupled differential equations for $\mathfrak{su}(4)$ Eqs. (\ref{eq:su4wn1})-(\ref{eq:su4wn15}), a numerical study for designing a Hamiltonian that realizes the gate 
will be needed. This and other kind of  problems will be subject of future work.

\section{Conclusions and perspectives}\label{CONclu}

In this work we have introduced \verb|symdyn|, a Python library designed to automate the WNM, overcoming challenges associated with high-dimensional quantum systems. Our library facilitates the computation of the factorized TEO, regardless of the ordering of exponentials in the factorization, and enables the calculation of simple and nested commutators, as well as similarity transformations. We have emphasized the crucial role of selecting an appropriate basis to maximize the method’s utility and have analytically recovered non-trivial results for low-order QLSs as examples.

Furthermore, we have demonstrated \verb|symdyn|'s robustness and efficiency for high-order QLSs by deriving the TEO differential equations for a system of two coupled time-dependent harmonic oscillators with time-dependent coupling. We have specialized our library for the specially unitary group \textit{SU}$(N)$ by providing a generic CWB for its associated algebra, making it applicable for arbitrary $N$.  We applied this framework to \textit{SU}$(2)$, \textit{SU}$(3)$ and \textit{SU}$(4)$, and used our results to obtain the Hadamard, $T$, and CNOT gates, completing the necessary gates for universal quantum computation. 

There is a growing interest in understanding the interplay between increasing the dimension of Lie algebras, linked to the number of components in a quantum system, and the ability to compute their time evolution \cite{Wiersema_2024, Surace_2024, Qvarfort_2025}. We expect our results for \textit{SU}$(N)$ to contribute to tackling these scaling challenges. Moreover, we expect \verb|Symdyn| to become a powerful tool for studying high-dimensional quantum systems and exploring new frontiers in quantum dynamics, optimal control, and quantum computation. 


Currently, we are developing \verb|symdyn II|, which leverages the composition property of Lie groups to provide explicit solutions for the TEO of QLSs. This extension aims to enhance the efficient resolution of the differential equations generated by the WNM, such as time-dependent coupled Riccati equations, also addressing their stability \cite{Cestnik_2024, Lohe_2025}. Additionally, we plan to expand our repository by incorporating new QLSs and analyzing their respective differential equations. 

In the near future, we aim to extend the WNM to the algebras of pseudo-orthogonal groups, which play a fundamental role in high-energy physics and interacting dynamical systems. These groups offer a symmetry-based approach to understanding quantum and classical correlations. In particular, the Lorentz group has widespread applications in high-energy physics, where quantum computation frameworks are being actively explored \cite{Di_Meglio_2024}.

\begin{acknowledgments}


The authors thank J. A. Helayël Neto, J. I. Rubiano-Murcia, and A. C. Duriez for their valuable and enriching discussions. D. M. T. also expresses gratitude to Sebastian Galliez Martínez and L. Pires for their assistance with the graphical representation of the Structure Tensor.

D. M. T. and A. Z. K. also acknowledge the support of Brazilian scientific and technological research agencies, including the Coordination for the Improvement of Higher Education Personnel (CAPES), the National Council for Scientific and Technological Development (CNPq), the Fundação de Amparo à Pesquisa do Estado de São Paulo (FAPESP),
grant 2021/06823-5 (A.Z.K) and the Fundação de Amparo à Pesquisa do Estado do Rio de Janeiro (FAPERJ), grants E-26/204.433/2021 (D.M.T.) and E-26/203.939/2024 (A.Z.K), for partial financial support.

\end{acknowledgments}


\bibliographystyle{unsrt}

\bibliography{Biblio}


\appendix

\onecolumngrid


\section{Wei-Norman Factorization} \label{appA}

Here, we show that an arbitrary element of a closed Lie group given in the unfactorized representation of Eq. (\ref{eq:unfacTEO}), can equivalently be expressed in the factorized representation of Eq. (\ref{eq:TEONcomp}) if $\boldsymbol{\xi}$ is invertible for all times. Recall that the functions relating the $\Lambda_{l}$ and  $\lambda_{l}$ coefficients are known as BCH-like  relations \cite{Barnett-1997}, and that we are considering the ordering given in Eq. (\ref{eq:TEONcompordered}). Following standard algebraic techniques, we re-define the operators in Eqs. (\ref{eq:unfacTEO}) and (\ref{eq:TEONcomp}), as the special case $\theta=1$ of 
\begin{align}
\hat{F}(\theta) =  \exp{\sum_{l=1}^{L}\theta\lambda_{l}\hat{g}_{l}} \, , 
\label{fdt}
\end{align}
and
\begin{align}
\hat{F} (\theta) = \prod_{l=1}^{L} e^{ \Lambda_{l}(\theta)\hat{g}_{l}} \, ,  
\label{fdt2}
\end{align}
respectively. Notice that we have suppressed the explicit time-dependence in the coefficients of the above equations once our results relating both representations are valid for (the same) any instant of time. The idea is to differentiate the above  representations of $\hat{F}$ with respect to $\theta$, rewrite them proportional to $\hat{F}$, and then impose their derivatives to be equal, establishing the set of differential equations that relates the set of coefficients in both representations.

  
The derivative of Eq. (\ref{fdt}) is direct, and given by
\begin{align}
\hat{F}'= &  \sum_{l=1}^{L}\lambda_{l}\hat{g}_{l}\hat{F} \,,
\label{fdt3}
\end{align}
where the prime indicates derivative with respect to $\theta$. On the other hand, the derivative of a factorized element with respect to any parameter is equivalent to the calculated in Eq. (\ref{eq:TEOderFin}). Therefore, for Eq. (\ref{fdt2}) we have
\begin{align}
\hat{F}' =  \sum_{l=1}^{L}\sum_{n=1}^{L}\Lambda'_{n}\xi_{nl}
\hat{g}_{l}\hat{F}  \, .
\label{eq:TEOderFinthe}
\end{align}
Using the linear independence of the generators, equaling Eqs. (\ref{fdt3}) and (\ref{eq:TEOderFinthe}) is obtained 
\begin{equation}
\lambda_{l}=\sum_{n=1}^{L}\Lambda'_{n}\xi_{nl} \, ,
\label{eq:diffeq}
\end{equation}
which is a set of coupled differential equations with the initial condition $\Lambda_{n}=0$ for $\theta=0$, and can be written in matrix notation as
\begin{equation}
\overbrace{\begin{bmatrix}
    \lambda_{1}     \\
    \lambda_{2}      \\
    \vdots		 \\
    \lambda_{L}   
\end{bmatrix}}^{
\mbox{$\boldsymbol{\lambda}$}
}
=
\overbrace{\begin{bmatrix}
     \xi_{11} & \xi_{21} & \dots  & \xi_{L1} \\
    \xi_{12} & \xi_{22} & \dots  & \xi_{L2} \\
    \vdots & \vdots & \ddots & \vdots \\
    \xi_{1L} & \xi_{2L} & \dots  & \xi_{LL}
\end{bmatrix}}^{
\mbox{$\boldsymbol{\xi}$}^{T}
}
\overbrace{\begin{bmatrix}
    \Lambda'_{1}     \\
    \Lambda'_{2}      \\
    \vdots		 \\
    \Lambda'_{L}   
\end{bmatrix}}^{
\mbox{$\boldsymbol{\Lambda}'$}
}\,\,.
\label{eq:diffeqmatrtind}
\end{equation}
The above system of differential equations is almost identical to such in Eq. (\ref{eq:diffeqmatr}) with the following differences: first and more important, there is not explicit dependence of $\lambda_{l}$ on the parameter $\theta$. Second, there is no imaginary unit at the right hand side of the equation. 
Accordingly, for a coupling matrix that is all time invertible we have
\begin{equation}
(\boldsymbol{\xi}^{T}) ^{-1} \boldsymbol{\lambda}= \boldsymbol{\Lambda}' \,,
\label{eq:Soldfactr}
\end{equation}
which is a system of block-decoupled nonlinear differential equations. Recall that once the system is solved, then it must be evaluated at $\theta=1$ to obtain the desired BCH-like relations.

To illustrate the above results, we next calculate the BCH-like relations for $\mathfrak{su}(1,1)$, $\mathfrak{su}(2)$, $\mathfrak{sl}(2)$ and $\mathfrak{so}(2,1)$ QLSs with the Lie products given in Table \ref{CommRe}.a. Using Eqs. (\ref{eq:xiMatrix3gen}) and  (\ref{eq:diffeqmatrtind}), we get
\begin{equation}
\begin{bmatrix}
    \lambda_{1}     \\
    \lambda_{2}      \\
    \lambda_{3}   
\end{bmatrix}
=
\begin{bmatrix}
    1 & -\upsilon\Lambda_{1}   & \epsilon\upsilon\frac{ (\Lambda_{1})^{2}}{e^{\upsilon\Lambda_{2}}} \\
    0 & 1  & -2\epsilon\frac{ \Lambda_{1}}{e^{\upsilon\Lambda_{2}}} \\
    0 & 0   & \frac{ 1}{e^{\upsilon\Lambda_{2}}}
\end{bmatrix}
\begin{bmatrix}
    \Lambda'_{1}     \\
    \Lambda'_{2}      \\
    \Lambda'_{3}   
\end{bmatrix}\,.
\label{eq:BCHqmatrQLS}
\end{equation}
As mentioned in Sec. \ref{coefficients}, once $\det\vert\boldsymbol{\xi}\vert=\exp(-\upsilon\Lambda_{2})\,$ one can invert $\boldsymbol{\xi}^{T}$ for all times, except for the limiting case $\Lambda_{2}\rightarrow\infty$ which nonetheless is tractable, as shown in Appendix~\ref{app:xgateCWB}. Then, multiplying by $(\boldsymbol{\xi}^{T}) ^{-1}$ from the left in the above equation, we obtain the system of block-decoupled differential equations:
\begin{align}
\label{eq:RiccatiInd}
&\Lambda'_{1}-\epsilon\upsilon\lambda_{3}(\Lambda_{1})^{2}- \upsilon\lambda_{2}\Lambda_{1} - \lambda_{1}=0 \, , \\
\label{eq:sollambdacTInd}
&\Lambda'_{2}-2\epsilon\lambda_{3}\Lambda_{1}- \lambda_{2} = 0 \, , \\
\label{eq:solgmabdacInd}
&\Lambda'_{3} - \lambda_{3} e^{\upsilon\Lambda_{2}} = 0 \,.
\end{align}
The first equation above represents four families of parameter-independent Riccati equations, each associated with one of the Lie algebras we are considering. Once this Riccati equation is solved, the other two equations can be integrated by quadrature using the initial conditions. 
The analytical solution of Eq.  (\ref{eq:RiccatiInd}) is well-known (see, for instance, the appendix of Ref. \cite{DMT-BJP-2019}). Moreover, the solution of the above equations is equivalent to the one in Refs. \cite{DMT-BCH-2020, DMT-PRA-2023} for the Lie algebras $\mathfrak{su}(1,1)$, $\mathfrak{su}(2)$ and $\mathfrak{so}(2,1)$, but now including the Lie algebra $\mathfrak{sl}(2)$. This solution is given as
\begin{align}
\Lambda_{1}=&\frac{2\lambda_{1} \sinh(\nu)}{2\nu \cosh(\nu)-\upsilon\lambda_{2}\sinh(\nu)}\,, \label{eq:lambda1}\\
\Lambda_{3}=&\frac{2\lambda_{3} \sinh(\nu)}{2\nu \cosh(\nu)-\upsilon\lambda_{2}\sinh(\nu)}\,,
\label{eq:sollambdac}
\end{align}
and
\begin{equation}
\Lambda_{2}=-\frac{2}{\upsilon}\ln{\left(\cosh(\nu)-\frac{\upsilon\lambda_{2}}{2\nu} \sinh(\nu)\right)}\, ,
\label{eq:lambdamenos}
\end{equation}
with
\begin{equation}
\nu^{2} = \left(\frac{\upsilon\lambda_{2}}{2}\right)^{2}-\epsilon\upsilon\lambda_{1}\lambda_{3} \, . 
\label{eq:nu2}
\end{equation}
The above equations are therefore BCH-like relations for the Lie algebras $\mathfrak{su}(1,1)$, $\mathfrak{su}(2)$, $\mathfrak{so}(2,1)$ and $\mathfrak{sl}(2)$. Recall that a different choice for the order of the generators, which means reordering the elements of the structure tensor in Table \ref{CommRe}.(a), will leads to a different set of functions for the above  coefficients.  
\section{Unitarity constraints for the $\mathfrak{su}(2)$ TEO } \label{Unit}

Here, we deduced the constraints to guarantee the unitarity of the TEO in Eq. (\ref{eq:TEONQubit}). From that equation,  it is direct that
\begin{align}
\hat{U}^{\dagger}=e^{-\frac{\Lambda_{2}^{*}}{2}}\begin{bmatrix}
    e^{\Lambda_{2}^{*}}+ \Lambda_{1}^{*}\Lambda_{3}^{*} & \Lambda_{3}^{*} 
   \\
     \Lambda_{1}^{*} & 1
\end{bmatrix}\, ,
\label{eq:TEONQubitCT}
\end{align}
where $\dagger$ represents the Hermitian conjugate and $*$ the complex conjugate. The unitarity condition implies $\hat{U}^{\dagger}\hat{U}=\hat{U}\hat{U}^{\dagger}=\hat{1\!\! 1}$. Accordingly, using Eqs. (\ref{eq:TEONQubit}) and (\ref{eq:TEONQubitCT}), it must be satisfied
\begin{align}
\begin{bmatrix}
    1 & 0 
   \\
     0 & 1
\end{bmatrix}=
e^{-\mathfrak{Re}(\Lambda_{2})}
\begin{bmatrix}
    \left|\kappa\right|^{2}+\left|\Lambda_{1}\right|^{2} & \kappa\Lambda_{3}^{*} + \Lambda_{1}
   \\
     \kappa^{*}\Lambda_{3} + \Lambda_{1}^{*} 
     & 1 + \left|\Lambda_{3}\right|^{2}
\end{bmatrix}\, ,
\label{eq:produn}
\end{align}
where $\kappa = e^{\Lambda_{2}}+ \Lambda_{1}\Lambda_{3}$, $\mathfrak{Re}(z)$ and $\mathfrak{Im}(z)$ represent the real and imaginary part of the complex number $z$, and the TEO coefficients are written in its polar form as $\Lambda_{1}=\left|\Lambda_{1}\right|e^{i \varphi_{1}}$, $\Lambda_{2}=\left|\Lambda_{2}\right|e^{i \varphi_{2}}$ and $\Lambda_{3}=\left|\Lambda_{3}\right|e^{i \varphi_{3}}$. Once $e^{-\mathfrak{Re}(\Lambda_{2})} \neq 0$, then Eq. (\ref{eq:produn}) implies the system of equations
\begin{align}
&e^{-\mathfrak{Re}(\Lambda_{2})}\left( \left|\kappa\right|^{2}+\left|\Lambda_{1}\right|^{2}\right)=1\,, \label{eq:cons1}\\
&e^{-\mathfrak{Re}(\Lambda_{2})}\left( 1+\left|\Lambda_{3}\right|^{2}\right)=1\,,
\label{eq:cons2}\\
&\kappa \left|\Lambda_{3}\right|e^{-i \varphi_{3}} + \left|\Lambda_{1}\right|e^{i \varphi_{1}}=0\,.
\label{eq:cons3}
\end{align}
Using Eq. (\ref{eq:cons3}) and its complex conjugated we first calculate $\left|\kappa\right|=\frac{\left|\Lambda_{1}\right|}{\left|\Lambda_{3}\right|}$, and then substituting in Eq. (\ref{eq:cons1}) we obtain
\begin{align}
&e^{-\mathfrak{Re}(\Lambda_{2})}\left( 1+\left|\Lambda_{3}\right|^{2}\right)=\frac{\left|\Lambda_{3}\right|^{2}}{\left|\Lambda_{1}\right|^{2}}\,. 
\label{eq:consp1}
\end{align}
From Eq. (\ref{eq:cons2}) and the above equation it is clear that $\left|\Lambda_{3}\right|=\left|\Lambda_{1}\right|$, which is the first constraint. Eq. (\ref{eq:cons2}) can be now rewritten as $e^{\mathfrak{Re}(\Lambda_{2})}=\left( 1+\left|\Lambda_{1}\right|^{2}\right)$, which is the second constraint. Now, substituting $\kappa$ explicitly in Eq. (\ref{eq:cons3}) and using the first constraint we obtain
\begin{align}
& e^{\Lambda_{2}}+ \left( 1+\left|\Lambda_{1}\right|^{2}\right)e^{i (\varphi_{1}+\varphi_{3})}=0\,.
\label{eq:cons3exp}
\end{align}
Finally, multiplying the above equation by $e^{-\mathfrak{Re}(\Lambda_{2})}$ and using the second constraint we get
\begin{align}
& e^{i\mathfrak{Im}(\Lambda_{2})} + e^{i (\varphi_{1}+\varphi_{3})}=0\,.
\label{eq:cons3fin}
\end{align}
The above equation is the third constraint, and it can be written as $\,\mathfrak{Im}(\Lambda_{2})=\varphi_{1}+\varphi_{3}+(2n+1)\pi\,$, with $n\in\mathbb{Z}$.
They are partially in agreement with those of Ref. \cite{DMT-PRA-2023} recalling that they considered $\ln{\Lambda_{2}}$ as the coefficient of $\hat{g}_{2}$ and our third constraint coincides with Eqs. (14) of that reference for $n=1$.

\section{X gate in the Cartan-Weyl basis}\label{app:xgateCWB}

Here we consider a different approach than such in Section \ref{OQgates} to obtain the $X$ gate
\begin{align}
     \hat{U}_x = \begin{bmatrix}
    0 & 1 \\
    1 & 0
    \end{bmatrix} \,.
\end{align}
The idea is to start from Eq.  (\ref{eq:TEONQubitfin2}), ask for the values that parameters $\left|\Lambda_{1}\right|$, $\varphi_{1}$ and $\varphi_{3}$ should take to get  the mentioned gate, and obtain the Hamiltonian coefficients that produces them using the WNM. This is a kind of an inverse problem.  Initially, to get a null diagonal, the only option is to have $\left|\Lambda_{1}\right|\rightarrow\infty$. Notice that this limit must also be taken as a step  to obtain the $Y$ gate, which is proportional to $\hat{\sigma}_2$. The $X$ gate is obtained, up to a global phase, if it is satisfied $\varphi_{1}=\varphi_{3}\equiv\varphi$, while to obtain the $Y$ gate it must be satisfied $\varphi_{1}=(2n+3/2)\pi$, with $n\in\mathbb{Z}$, and $\varphi_{3}=(2m+1/2)\pi$, with $m\in\mathbb{Z}$. From the first two unitarity constraints we have that $\vert\Lambda_{3}\vert\rightarrow\infty\,$ and 
$e^{\mathfrak{Re}(\Lambda_{2})}\rightarrow\infty$, so it is evident that to obtaining such a gates it is necessary to take the limit $\Lambda_{2}\rightarrow\infty$, in which the coupling matrix cannot be inverted (see Eq.  (\ref{eq:diffeqmatrQLS})).  Notice that the third unitarity constraint implies for the $X$ gate that $e^{i\mathfrak{Im}(\Lambda_{2})}=-e^{i\varphi}\,$.

It is well known that the $X$ gate  can be obtained from a Hamiltonian proportional to $\hat{\sigma}_1$. Once our Hamiltonian in Eq. (\ref{eq:rabiHamiltonian}) is transformed to the CWB, the above implies to put $\eta_2 =0$. We also  define $\eta_1=\eta_3\equiv\eta$. 
This yields
\begin{align}
\hat{H} = \eta(\hat{g}_{1}+\hat{g}_{3})\,.
\label{eq:rabiHamiltonianinx}
\end{align}
As the above Hamiltonian is time-independent, it is direct that
\begin{align}
\hat{U}= \exp{\lambda(\hat{g}_{1}+\hat{g}_{3})}\,,
\label{eq:rabiHamiltonianinx}
\end{align}
where $\lambda=i\eta t$. Accordingly, the problem has been reduced to a factorization problem, for which we can use Eqs. (\ref{eq:lambda1})-(\ref{eq:nu2}) with $\epsilon=-1$ and $\upsilon=1$, $\lambda_2=0$ and $\lambda_1=\lambda_3=\lambda$,  yielding $\Lambda_{1}=\Lambda_{3}=\tanh{(\lambda)}\,$ and $\Lambda_{2}=-2\ln{\cosh{(\lambda)}}$. In terms of the Hamiltonian coefficient, we obtain $\Lambda_{1}=\Lambda_{3}=-i\tan{(\eta t)}\,$ and $\Lambda_{2}=-2\ln{\cos{(\eta t)}}$. With these solutions, the TEO is explicitly
\begin{align}
\hat{U}(t) = \begin{bmatrix}
\cos(\eta t) & -i\sin (\eta t)\\
-i\sin(\eta t) & \cos (\eta t)
\end{bmatrix}.
\end{align}
At $t=t^\star:=(n+1/2)\pi/\eta\,$, with $n\in\mathbb{Z}$, the TEO $\hat{U}(t^\star)$ is the X gate up to a global phase of $ i^{(2n-1)}$.

\section{Time-dependent coupled oscillators} \label{appD}

In this appendix, we provide some details for the problem of 
two time-dependent harmonic oscillators coupled with a time-dependent coupling of Section \ref{couposc}. As we mentioned, the effective Hamiltonian of the system can be written as in Eq. (\ref{eq:HamCoup}), where the generators are bilinear products of ladder operators of each mode, and the coefficients are functions of the physical parameters of the system \textit{i.e.}, the mirror position and its derivative, the frequency of the modes, and the coupling parameters being elements of a tensor representing all the intermode interactions \cite{Ramos_2021}. Note that other types of interaction between modes can give rise to underlying non-closed Lie algebras \cite{Edward_Bruschi_2024}. To construct the Structure Tensor, we shall write explicitly the generators of the system. Let $\hat{a}_{j}$ and $\hat{a}^{\dagger}_{j}$ be ladder operators of the mode $j$, with $j=1,2$. Accordingly, they satisfy 
\begin{align}
[\hat{a}_{j},\hat{a}_{k}]=0  \,\,\, \mbox{and} \,\,\,
[\hat{a}_{j},\hat{a}^{\dagger}_{k}]=\delta_{jk}\hat{1\!\! 1} \,.
\label{eq:LadderAlg}
\end{align}
The eleven generators are defined as 
\begin{align}
\hat{g}_{1} =&  \left. \hat{a}^{\dagger}_{1}\right. ^{2}, 
\,\,\, 
\hat{g}_{2} = \left. \hat{a}^{\dagger}_{2}\right. ^{2}, 
\,\,\,
\hat{g}_{3} = \hat{a}^{\dagger}_{1} \hat{a}^{\dagger}_{2},
\,\,\,
\hat{g}_{4} = \hat{a}^{\dagger}_{1} \hat{a}_{2}, 
 \nonumber\\[2pt]
\hat{g}_{5} =&  \hat{a}_{1} \hat{a}^{\dagger}_{2}, 
\,\,\, 
\hat{g}_{6} = \hat{a}^{\dagger}_{1} \hat{a}_{1}, 
\,\,\,
\hat{g}_{7} = \hat{a}^{\dagger}_{2} \hat{a}_{2},
\,\,\,
\hat{g}_{8} = \hat{a}_{1}^{2}, 
\nonumber\\[2pt] 
\hat{g}_{9} =&  \hat{a}_{2}^{2}, 
\,\,\, 
\hat{g}_{10} = \hat{a}_{1} \hat{a}_{2}, 
\,\,\,
\hat{g}_{11} = \hat{1\!\! 1} \, .
\label{eq:GenCoup}
\end{align}
using Eqs. (\ref{eq:LadderAlg}), we calculate the commutators between the above generators and build the correspondent structure table of the system, Table \ref{CommRecoupled}, displayed at the end of this appendix.  From it can be noted that $\hat{g}_{6}$ and $\hat{g}_{7}$ are members of the Cartan subalgebra, and with exception of $\hat{g}_{11}$, all other generators belong to the root spaces. Recall that the determinant of the correspondent coupling matrix involves the exponential of $\hat{g}_{6}$ and $\hat{g}_{7}$.  

\begin{center}
\begin{table}[htb]
\centering
\begin{tabular}{|l|c|c|c|c|c|c|c|c|c|c|c|}
 \hline
  & $\hat{g}_{1}$ & $\hat{g}_{2}$ & $\hat{g}_{3}$  & $\hat{g}_{4}$  & $\hat{g}_{5}$  & $\hat{g}_{6}$  & $\hat{g}_{7}$  & $\hat{g}_{8}$  & $\hat{g}_{9}$  & $\hat{g}_{10}$  & $\hat{g}_{11}$\\
 \hline
$\hat{g}_{1}$  & $0$  & $0$ & $0$ & $0$ & $-2\hat{g}_{3}$ & $-2\hat{g}_{1}$ & $0$ & $-4\hat{g}_{6}-2\hat{g}_{11}$ & $0$ & $-2\hat{g}_{4}$ & $0$  \\
\hline
$\hat{g}_{2}$ & * & $0$ & $0$ & $-2\hat{g}_{3}$ & $0$ & $0$ & $-2\hat{g}_{2}$ & $0$ & $-4\hat{g}_{7}-2\hat{g}_{11}$ & $-2\hat{g}_{5}$ & $0$  \\
\hline
$\hat{g}_{3}$ & * & * & $0$ & $-\hat{g}_{1}$ & $-\hat{g}_{2}$ & $-\hat{g}_{3}$ & $-\hat{g}_{3}$ & $-2\hat{g}_{5}$ & $-2\hat{g}_{4}$ & $-\hat{g}_{6}-\hat{g}_{7}-\hat{g}_{11}$ & $0$  \\
\hline
$\hat{g}_{4}$ & * & * & * & $0$ & $\hat{g}_{6}-\hat{g}_{7}$ & $-\hat{g}_{4}$ & $\hat{g}_{4}$ & $-2\hat{g}_{10}$ & $0$ & $-\hat{g}_{9}$ & $0$  \\
\hline
$\hat{g}_{5}$ & * & * & * & * & $0$ & $\hat{g}_{5}$ & $-\hat{g}_{5}$ & $0$ & $-2\hat{g}_{10}$ & $-\hat{g}_{8}$ & $0$  \\
\hline
$\hat{g}_{6}$ & * & * & * & * & * & $0$ & $0$ & $-2\hat{g}_{8}$ & $0$ & $-\hat{g}_{10}$ & $0$  \\
\hline
$\hat{g}_{7}$ & * & * & * & * & * & * & $0$ & $0$ & $-2\hat{g}_{9}$ & $-\hat{g}_{10}$ & $0$  \\
\hline
$\hat{g}_{8}$ & * & * & * & * & * & * & * & $0$ & $0$ & $0$ & $0$  \\
\hline
$\hat{g}_{9}$ & * & * & * & * & * & * & * & * & $0$ & $0$ & $0$  \\
\hline
$\hat{g}_{10}$ & * & * & * & * & * & * & * & * & * & $0$ & $0$  \\
\hline
$\hat{g}_{11}$ & * & * & * & * & * & * & * & * & * & * & $0$  \\
\hline
\end{tabular}  
\label{CommRecoupled}
\caption{Structure table for a system of two coupled time-dependent harmonic oscillators. The symbol $*$ is used here to indicate that such entries are known once this table is anti-symmetric.}
\end{table}
\end{center}

\section{key results of the WNM for $\mathfrak{su}(4)$}\label{sec:su4B}

Here, we use \verb|symdyn| to calculate some key results of the WNM for $\mathfrak{su}(4)$.  Initially, taking $N=4$ in Eq. (\ref{eq:gensunordering}), we obtain the generic CWB for $\mathfrak{su}(3)$
\begin{align}
&\hat{g}_{1}=
 \begin{bmatrix}
   0 & 1 & 0 & 0 \\
	0 & 0 & 0 & 0 \\
	 0 & 0 & 0 & 0 \\
	  0 & 0 & 0 & 0
\end{bmatrix},  \,\, 
 \cdots, \,\,
\hat{g}_{6}=
\begin{bmatrix}
   0 & 0 & 0 & 0 \\
	0 & 0 & 0 & 0 \\
	 0 & 0 & 0 & 1 \\
	  0 & 0 & 0 & 0
\end{bmatrix} , \,\,
\hat{g}_{7}=
\begin{bmatrix}
   1 & 0 & 0 & 0 \\
	0 & -1 & 0 & 0 \\
	 0 & 0 & 0 & 0 \\
	  0 & 0 & 0 & 0
\end{bmatrix},  \,\, 
\cdots , \,\,
\nonumber \\
&\hat{g}_{9}=
\begin{bmatrix}
   0 & 0 & 0 & 0 \\
	0 & 0 & 0 & 0 \\
	 0 & 0 & 1 & 0 \\
	  0 & 0 & 0 & -1
\end{bmatrix}, \,\,
\hat{g}_{10}=
\begin{bmatrix}
   0 & 0 & 0 & 0 \\
	1 & 0 & 0 & 0 \\
	 0 & 0 & 0 & 0 \\
	  0 & 0 & 0 & 0
\end{bmatrix},  \,\, 
 \cdots, \,\,
\hat{g}_{15}=
\begin{bmatrix}
   0 & 0 & 0 & 0 \\
	0 & 0 & 0 & 0 \\
	 0 & 0 & 0 & 0 \\
	  0 & 0 & 1 & 0
\end{bmatrix} .  
\label{eq:CWBsu3}
\end{align}
The TEO now reads
\begin{align}
\hat{U}(t) = \prod_{l=1}^{15} e^{ \Lambda_{l}(t)\hat{g}_{l}} \, .
\label{eq:TEOSU3c}
\end{align}

We can continue to ask \verb|symdyn| for the determinant of the coupling matrix of the system,  which results to be  $\det\vert\boldsymbol{\xi}\vert = \exp{-2(\Lambda_{7}+\Lambda_{8}+\Lambda_{9})}\,$, involving the coefficients of the generators belonging to the Cartan subalgebra. Accordingly, we can ask \verb|symdyn| for the system of block-decoupled differential equations, which returns
\begin{small}
\begin{align}
\label{eq:su4wn1}
i \dot{\Lambda}_1 = & - \Lambda_1^2 \eta_{10} - \Lambda_1 (\Lambda_2 \eta_{11} + \Lambda_3 \eta_{12}) + \Lambda_1 (2 \eta_7 - \eta_8)  - \Lambda_2 \eta_{13} - \Lambda_3 \eta_{14}+ \eta_1  \, ,\\
\label{eq:su4wn2}
i \dot{\Lambda}_2 = & - \Lambda_2^2 \eta_{11} - \Lambda_2 (\Lambda_1 \eta_{10} + \Lambda_3 \eta_{12}) + \Lambda_2 (\eta_7 + \eta_8 - \eta_9) - \Lambda_3 \eta_{15} - \Lambda_1 \eta_4 + \eta_2\,,\\
\label{eq:su4wn3}
i \dot{\Lambda}_3 = & - \Lambda_3^2 \eta_{12} - \Lambda_3 (\Lambda_1 \eta_{10} + \Lambda_2 \eta_{11} ) + \Lambda_3 (\eta_7 + \eta_9) - \Lambda_1 \eta_5 - \Lambda_2 \eta_6  + \eta_3\,,\\
\label{eq:su4wn4}
i \dot{\Lambda}_4 = & -\Lambda_4^2 (\Lambda_1  \eta_{11}  + \eta_{13}) - \Lambda_4 \Lambda_5 (\Lambda_1 \eta_{12} + \eta_{14} ) + \Lambda_4 (\Lambda_1 \eta_{10} - \Lambda_2 \eta_{11} -\eta_7 + 2 \eta_8 - \eta_9) 
- \Lambda_5 (\Lambda_2 \eta_{12} + \eta_{15}) + \Lambda_2 \eta_{10} + \eta_4 \,,\\
\label{eq:su4wn5}
i \dot{\Lambda}_5 = &  -  \Lambda_5^2 (\Lambda_1 \eta_{12} + \eta_{14}) - \Lambda_4 \Lambda_5  (\Lambda_1\eta_{11} + \eta_{13})  +  \Lambda_5 (\Lambda_1 \eta_{10}  - \Lambda_3 \eta_{12}  - \eta_7 + \eta_8 + \eta_9) -  \Lambda_4 (\Lambda_3 \eta_{11} + \eta_6) + \Lambda_3 \eta_{10} + \eta_5 \,,\\
\label{eq:su4wn6}
i \dot{\Lambda}_6 = & - \Lambda_6^2 ( \Lambda_2 \eta_{12} + \eta_{15} + \Lambda_4(\Lambda_1  \eta_{12} + \eta_{14})) -  \Lambda_5 \Lambda_6 (\Lambda_1\eta_{12} + \eta_{14}) + \Lambda_6 ( \Lambda_2 \eta_{11} +\Lambda_4 (\Lambda_1 \eta_{11}  + \eta_{13}) - \Lambda_3 \eta_{12} - \eta_8 + 2 \eta_9) + \nonumber\\ 
& + \Lambda_3 \eta_{11}  + \Lambda_5 (\Lambda_1 \eta_{11} +\eta_{13}) + \eta_6\,,\\
\label{eq:su4wn7}
i \dot{\Lambda}_7 = & - \Lambda_1 \eta_{10} - \Lambda_2 \eta_{11} - \Lambda_3 \eta_{12} +  \eta_7\,,\\
\label{eq:su4wn8}
i \dot{\Lambda}_8 = &   - \Lambda_2 \eta_{11} - \Lambda_3 \eta_{12} -  \Lambda_4 (\Lambda_1 \eta_{11} + \eta_{13}) - \Lambda_5 (\Lambda_1  \eta_{12} + \eta_{14}) + \eta_8\,,\\
\label{eq:su4wn9}
i \dot{\Lambda}_9 = &    - \Lambda_3 \eta_{12}  -  \Lambda_5 (\Lambda_1 \eta_{12} + \eta_{14}) - \Lambda_6 (\Lambda_2 \eta_{12} + \eta_{15} + \Lambda_4(\Lambda_1 \eta_{12} + \eta_{14})) + \eta_9\,,\\
\label{eq:su4wn10}
i \dot{\Lambda}_{10} = & (\eta_{10}  - \Lambda_4 \eta_{11} - \Lambda_5 \eta_{12}) e^{(2 \Lambda_7 - \Lambda_8)} \,,\\
\label{eq:su4wn11}
i \dot{\Lambda}_{11} = & \Lambda_{10} (\Lambda_1 \eta_{11} + \eta_{13} -  \Lambda_6 (\Lambda_1 \eta_{12} +\eta_{14})) e^{(- \Lambda_7 + 2 \Lambda_8 - \Lambda_9)} + (\eta_{11} - \Lambda_6 \eta_{12}) e^{(\Lambda_7 + \Lambda_8 - \Lambda_9)} \,,\\
\label{eq:su4wn12}
i \dot{\Lambda}_{12} = &  \Lambda_{10} (\Lambda_1 \eta_{12}  + \eta_{14}) e^{(- \Lambda_7 + \Lambda_8 + \Lambda_9)} + \Lambda_{11} (\Lambda_2 \eta_{12}  + \eta_{15} + \Lambda_4 (\Lambda_1  \eta_{12}  + \eta_{14})) e^{(- \Lambda_8 + 2 \Lambda_9)} + \eta_{12} e^{(\Lambda_7 + \Lambda_9)} \,,\\
\label{eq:su4wn13}
i \dot{\Lambda}_{13} = & (\Lambda_1 \eta_{11}  + \eta_{13} -  \Lambda_6 (\Lambda_1 \eta_{12}+ \eta_{14})) e^{(- \Lambda_7 + 2 \Lambda_8 - \Lambda_9)}\,,\\
\label{eq:su4wn14}
i \dot{\Lambda}_{14} = &  \Lambda_{13} ( \Lambda_2 \eta_{12}  + \eta_{15} + \Lambda_4 (\Lambda_1  \eta_{12} + \eta_{14})) e^{(- \Lambda_8 + 2 \Lambda_9)} + (\Lambda_1 \eta_{12}  + \eta_{14}) e^{(- \Lambda_7 + \Lambda_8 + \Lambda_9)} \,,\\
\label{eq:su4wn15}
i \dot{\Lambda}_{15} = & ( \Lambda_2 \eta_{12}+ \eta_{15}  + \Lambda_4 (\Lambda_1\eta_{12} +  \eta_{14})) e^{- \Lambda_8 + 2 \Lambda_9} 
\, . 
\end{align}
\end{small}As can it be seen, the first three equations (the same number as the rank of this algebra) form a block of coupled Riccati equations, which, once solved, allow us to solve the fourth and fifth equations, which are themselves coupled Riccati equations. Then, we can move on to solve the sixth equation, which is a Riccati equation, and the rest can be solved by quadrature. As mentioned, this hierarchy of Riccati equations is expected. 

The explicit form of the TEO for this algebra is given as
%
\begin{align}
\hat{U}=\begin{bmatrix}
\Lambda_{1} \Lambda_{10} \varrho_{3} + \Lambda_{11} \varrho_{4} + \Lambda_{13} \varrho_{6}e^{-\Lambda_{9}} + e^{\Lambda_{7}} & \Lambda_{12} \varrho_{4} + \Lambda_{14} \varrho_{6}e^{-\Lambda_{9}} + \Lambda_{1} \varrho_{3} & \Lambda_{15} \varrho_{6}e^{-\Lambda_{9}} + \varrho_{4} & \varrho_{6}e^{-\Lambda_{9}} \\
\Lambda_{10} \varrho_{3} + \Lambda_{11} \Lambda_{4} \varrho_{5} + \Lambda_{13} \varrho_{7}e^{-\Lambda_{9}} & \Lambda_{12} \Lambda_{4} \varrho_{5} + \Lambda_{14} \varrho_{7}e^{-\Lambda_{9}} + \varrho_{3} & \Lambda_{15} \varrho_{7}e^{-\Lambda_{9}} + \Lambda_{4} \varrho_{5} & \varrho_{7}e^{-\Lambda_{9}} \\
\Lambda_{11} \varrho_{5} + \Lambda_{13} \Lambda_{6} e^{-\Lambda_{9}} & \Lambda_{12} \varrho_{5} + \Lambda_{14} \Lambda_{6} e^{-\Lambda_{9}} & \Lambda_{15} \Lambda_{6} e^{-\Lambda_{9}} + \varrho_{5} & \Lambda_{6} e^{-\Lambda_{9}} \\
\Lambda_{13} e^{-\Lambda_{9}} & \Lambda_{14} e^{-\Lambda_{9}} & \Lambda_{15} e^{-\Lambda_{9}} & e^{-\Lambda_{9}}
\end{bmatrix}\, ,
\label{eq:TEONsu4CWB}
\end{align}
where $\varrho_{3}=e^{(\Lambda_8-\Lambda_7)}$, $\varrho_{4}=(\Lambda_1 \Lambda_4 + \Lambda_2)\varrho_{5}\,$ with $\varrho_{5}=e^{(-\Lambda_8+\Lambda_9)}$, $\varrho_{6}=\Lambda_1 \Lambda_5 + \Lambda_3 + \Lambda_6 (\Lambda_1 \Lambda_4 + \Lambda_2)$, and $\varrho_{7}=\Lambda_4 \Lambda_6 + \Lambda_5$. 
A comparison between the above TEO and the CNOT gate in Eq. (\ref{eq:CNOTM}),  
\begin{align}
    \hat{U}_{CNOT} = \begin{bmatrix}
   1 & 0 & 0 & 0 \\
	0 & 1 & 0 & 0 \\
	 0 & 0 & 0 & 1 \\
	  0 & 0 & 1 & 0
\end{bmatrix} \,,
\end{align}
yields some direct constraints on the components of $\hat{U}$. From $\hat{U}_{44}$, it is evident that  obtaining the CNOT gate requires taking the limit $\Lambda_{9}\rightarrow\infty$. To achieve the desired  values for $\hat{U}_{34}$ and $\hat{U}_{43}$ in this  limit, the conditions $\Lambda_{6}= e^{\Lambda_{9}}$ and $\Lambda_{15}= e^{\Lambda_{9}}$ appear necessary. However, recalling that the unitarity constraints has not yet been applied to the TEO — and that the number of real variables currently exceeds the required $16$— we introduce a phase $\phi_n$ through the definitions  $\Lambda_{6}= e^{\Lambda_{9}}e^{i\phi_n}$ and $\Lambda_{15}= e^{\Lambda_{9}}e^{i\phi_n}$, allowing for an additional flexibility and that will be determined later. To ensure the necessary zeros in the last row and column of the TEO as $\Lambda_{9}\rightarrow\infty$, it suffices for the complex functions $\Lambda_{13}$, $\Lambda_{14}$, $\varrho_{6}$, and $\varrho_{7}$, to remain bounded. Additionally, achieving $e^{i\phi_n}$ as a global phase requires $\hat{U}_{11}=e^{i\phi_n}$ and $\hat{U}_{22}=e^{i\phi_n}$. Accordingly, we now derive the equations that arise from equating $\hat{U}$ and $\hat{U}_{CNOT}$ using the definitions $\Lambda_{6}= e^{\Lambda_{9}}e^{i\phi_n}$ and $\Lambda_{15}= e^{\Lambda_{9}}e^{i\phi_n}$, and expressing the results in terms of $\varrho_{6}$, and $\varrho_{7}$ 
\begin{align}
\label{eq:1U11}
& \Lambda_{10} \Lambda_1 e^{(-\Lambda_7 + \Lambda_8)} + \Lambda_{11} (\Lambda_1 \Lambda_4 + \Lambda_2) e^{(-\Lambda_8 + \Lambda_9)} + \Lambda_{13} \varrho_{6} e^{-\Lambda_9} + e^{\Lambda_7} = e^{i\phi_n}, \\
\label{eq:1U12}
& \Lambda_{12} (\Lambda_1 \Lambda_4 + \Lambda_2) e^{(-\Lambda_8 + \Lambda_9)} + \Lambda_{14} \varrho_{6} e^{-\Lambda_9} + \Lambda_1 e^{(-\Lambda_7 + \Lambda_8)} = 0, \\
\label{eq:1U13}
& \varrho_{6} e^{i\phi_n} + (\Lambda_1 \Lambda_4 + \Lambda_2) e^{(-\Lambda_8 + \Lambda_9)} = 0, \\
\label{eq:1U14}
& \varrho_{6} = \Lambda_1 \Lambda_5 + \Lambda_3 + (e^{\Lambda_9}e^{i\phi_n})(\Lambda_1 \Lambda_4 + \Lambda_2), \quad \text{and equivalently:} \quad \varrho_{6} = \Lambda_2 (e^{\Lambda_9}e^{i\phi_n}) + \Lambda_3 + \Lambda_1  \varrho_{7}, \\
\label{eq:1U21}
& \Lambda_{10} e^{(-\Lambda_7 + \Lambda_8)} + \Lambda_{11} \Lambda_4 e^{(-\Lambda_8 + \Lambda_9)} + \Lambda_{13} \varrho_{7} e^{-\Lambda_9} = 0, \\
\label{eq:1U22}
& \Lambda_{12} \Lambda_4 e^{(-\Lambda_8 + \Lambda_9)} + \Lambda_{14} \varrho_{7} e^{-\Lambda_9} + e^{(-\Lambda_7 + \Lambda_8)} = e^{i\phi_n}, \\
\label{eq:1U23}
& \varrho_{7} e^{i\phi_n} + \Lambda_4 e^{(-\Lambda_8 + \Lambda_9)} = 0, \\
\label{eq:1U24}
& \varrho_{7} = \Lambda_4 (e^{\Lambda_9}e^{i\phi_n}) + \Lambda_5, \\
\label{eq:1U31}
& \Lambda_{11} e^{(-\Lambda_8 + \Lambda_9)} + \Lambda_{13} e^{i\phi_n} = 0, \\
\label{eq:1U32}
& \Lambda_{12} e^{(-\Lambda_8 + \Lambda_9)} + \Lambda_{14} e^{i\phi_n} = 0, \\
\label{eq:1U33}
& e^{\Lambda_9}e^{2i\phi_n} + e^{(-\Lambda_8 + \Lambda_9)} = 0, \\
\label{eq:1U34}
& \Lambda_{6}= e^{\Lambda_{9}}e^{i\phi_n}, \\
\label{eq:1U41}
& \Lambda_{13} e^{-\Lambda_9} = 0, \\
\label{eq:1U42}
& \Lambda_{14} e^{-\Lambda_9} = 0, \\
\label{eq:1U43}
& \Lambda_{15}= e^{\Lambda_{9}}e^{i\phi_n},
\end{align}
where we used Eqs. (\ref{eq:1U14}), (\ref{eq:1U24}), (\ref{eq:1U34}), and (\ref{eq:1U43}), to explicitly express the definitions of $\varrho_{6}$, $\varrho_{7}$, $\Lambda_{6}$, and $\Lambda_{15}$, respectively. The equation for $\hat{U}_{44}$ has been suppressed, as it becomes equivalent to taking the limit $\Lambda_{9}\rightarrow\infty$ at the end of the calculations. From Eq. (\ref{eq:1U33}) we have that $\Lambda_8=-2i\phi_n +i\pi$. Accordingly, the equations involving this term, Eqs. (\ref{eq:1U11}), (\ref{eq:1U12}), (\ref{eq:1U13}), (\ref{eq:1U21}), (\ref{eq:1U22}), (\ref{eq:1U23}), (\ref{eq:1U31}) and (\ref{eq:1U32}), are modified to 
\begin{align}
\label{eq:2U11}
& -\Lambda_{10} \Lambda_1 e^{(-\Lambda_7 - 2i\phi_n)} - \Lambda_{11} (\Lambda_1 \Lambda_4 + \Lambda_2)e^{(2i\phi_n + \Lambda_9)} + \Lambda_{13} \varrho_6 e^{-\Lambda_9} + e^{\Lambda_7} = e^{i\phi_n}, \\
\label{eq:2U12}
& -\Lambda_{12} (\Lambda_1 \Lambda_4 + \Lambda_2)e^{(2i\phi_n + \Lambda_9)} + \Lambda_{14} \varrho_6 e^{-\Lambda_9} - \Lambda_1 e^{(-\Lambda_7 - 2i\phi_n)} = 0, \\
\label{eq:2U13}
& \varrho_6 e^{i\phi_n} - (\Lambda_1 \Lambda_4 + \Lambda_2)e^{(2i\phi_n + \Lambda_9)} = 0, \\
\label{eq:2U21}
& -\Lambda_{10} e^{(-\Lambda_7 - 2i\phi_n)} - \Lambda_{11} \Lambda_4 e^{(2i\phi_n + \Lambda_9)} + \Lambda_{13} \varrho_7 e^{-\Lambda_9} = 0, \\
\label{eq:2U22}
& -\Lambda_{12} \Lambda_4 e^{(2i\phi_n + \Lambda_9)} + \Lambda_{14} \varrho_7 e^{-\Lambda_9} - e^{(-\Lambda_7 - 2i\phi_n)} = e^{i\phi_n}, \\
\label{eq:2U23}
& \varrho_7 e^{i\phi_n} - \Lambda_4 e^{(2i\phi_n + \Lambda_9)} = 0, \\
\label{eq:2U31}
& -\Lambda_{11} e^{(2i\phi_n + \Lambda_9)} + \Lambda_{13} e^{i\phi_n} = 0, \\
\label{eq:2U32}
& -\Lambda_{12} e^{(2i\phi_n + \Lambda_9)} + \Lambda_{14} e^{i\phi_n} = 0, \\
\label{eq:2U33}
& \Lambda_8=-2i\phi_n +i\pi, 
\end{align}
where we wrote in Eq. (\ref{eq:2U33}) the definition of $\Lambda_8$. Multiplying Eqs. (\ref{eq:2U31}) and (\ref{eq:2U32}) by $e^{(-2i\phi_n+\Lambda_9)}$, and recalling that  we are considering $\Lambda_{13}$ and $\Lambda_{14}$ as free parameters, we obtain 
\begin{align}
\label{eq:pas1}
& \Lambda_{11} = \Lambda_{13}e^{(-i\phi_n-\Lambda_9)}, \\
\label{eq:pas2}
& \Lambda_{12} = \Lambda_{14}e^{(-i\phi_n-\Lambda_9)}.
\end{align}
From Eqs. (\ref{eq:2U23}) and (\ref{eq:1U24}) it is clear that $\Lambda_5=0$. From Eqs. (\ref{eq:2U13}) and (\ref{eq:1U14}) we have that $\Lambda_1\Lambda_5+\Lambda_3=0$, so $\Lambda_{3}=0$. Using explicitly $\varrho_6$ and $\varrho_7$, together with the above results, we can rewrite equations (\ref{eq:2U11})-(\ref{eq:2U23}) as
\begin{align}
\label{eq:3U11}
& -\Lambda_{10} \Lambda_1 e^{(-\Lambda_7 - 2i\phi_n)} + e^{\Lambda_7} = e^{i\phi_n}, \\
\label{eq:3U12}
& - \Lambda_1 e^{(-\Lambda_7 - 2i\phi_n)} = 0, \\
\label{eq:3U14}
& \varrho_6 = (e^{\Lambda_9}e^{i\phi_n})(\Lambda_1 \Lambda_4 + \Lambda_2), \\
\label{eq:3U21}
& -\Lambda_{10} e^{(-\Lambda_7 - 2i\phi_n)} = 0, \\
\label{eq:3U22}
&  - e^{(-\Lambda_7 - 2i\phi_n)} = e^{i\phi_n}, \\
\label{eq:3U24}
& \varrho_7 = \Lambda_4 (e^{\Lambda_9}e^{i\phi_n}), \\
\label{eq:3U13}
& \Lambda_3 = 0, \\
\label{eq:3U23}
& \Lambda_5 = 0, 
\end{align}
where we used the last two equations to explicitly write the previous findings. From Eqs. (\ref{eq:3U12}) and (\ref{eq:3U21}), we find that $\Lambda_1=0$ and $\Lambda_{10}=0$. Substituting these results, Eq. (\ref{eq:3U11}) yields $e^{\Lambda_7} = e^{i\phi_n}$, while Eq. (\ref{eq:3U22}) gives  $e^{\Lambda_7} = e^{-3i\phi_n+i(2n+1)\pi}$, where $n\in\mathbb{Z}$. By equating these expressions, we obtain $\phi_n=\frac{(2n+1)\pi}{4}$. Furthermore, to ensure $\hat{U}_{14}=0$ and $\hat{U}_{24}=0$ in the limit $\Lambda_{9}\rightarrow\infty$, conditions $\Lambda_{2}=\Lambda_{4}=0$ must be met. 
In summary, $\Lambda_{6}= \Lambda_{15}=e^{\Lambda_{9}+i\phi_n}$, $\Lambda_{7}=i\phi_n$, $\Lambda_{8}=i\pi(\frac{1}{2}-n)$, $\Lambda_{9}\rightarrow\infty$, $\Lambda_{11}= \Lambda_{13}e^{(-i\phi_n-\Lambda_9)}$, $\Lambda_{12}= \Lambda_{14}e^{(-i\phi_n-\Lambda_9)}$, with $\phi_n=\frac{(2n+1)\pi}{4}$, while $\Lambda_{1}=\Lambda_{2}=\Lambda_{3}=\Lambda_{4}=\Lambda_{5}=\Lambda_{10}=0$. $\Lambda_{13}$ and $\Lambda_{14}$ remain as free complex parameters.

\end{document}